\documentclass[runningheads]{llncs}
\usepackage{adjustbox,hyperref}
\usepackage{graphicx}
\usepackage{amsmath}
\usepackage{amssymb}
\usepackage{etoolbox}
\usepackage[misc,geometry]{ifsym}
\usepackage{xspace}
\usepackage{proof,url}
\usepackage{todonotes}
\usepackage{makecell}
\usepackage{mathtools}
\usepackage{multirow}
\usepackage{longtable}
\usepackage{algorithm}
\usepackage{vampire_induction}

\newbool{shortversion}
\boolfalse{shortversion}

\makeatletter
\newcommand{\leqnomode}{\tagsleft@true\let\veqno\@@leqno}
\newcommand{\reqnomode}{\tagsleft@false\let\veqno\@@eqno}
\makeatother

\def\orcidID#1{\href{http://orcid.org/#1}{\raisebox{-1.25pt}{\includegraphics{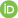}}}}

\newcommand{\reserved}[1]{\textbf{\underline{#1}}} 
\begin{document}
\title{Getting Saturated with Induction}
%
%
\author{M\'arton Hajdu\inst{1}\orcidID{0000-0002-8273-2613} \and
Petra Hozzov\'a\inst{1}\orcidID{0000-0003-0845-5811}(\Letter) \and
Laura Kovács\inst{1}\orcidID{0000-0002-8299-2714} \and
Giles Reger\inst{2}\orcidID{0000-0001-6353-952X} \and
Andrei Voronkov\inst{1,2,3} \\
\email{\{marton.hajdu, petra.hozzova\}@tuwien.ac.at}}
 \authorrunning{Hajdu, Hozzov\'a et al.}
%
\institute{TU Wien \and
University of Manchester \and EasyChair
}
\maketitle              
\begin{abstract}
Induction in saturation-based first-order theorem proving is a new exciting
direction in the automation of inductive reasoning. 
In this paper we survey our work on integrating induction directly  into the saturation-based proof search framework of first-order theorem proving. 
We describe our induction inference rules proving properties with  inductively defined datatypes and integers. We also present  additional reasoning heuristics 
for strengthening inductive reasoning, as well as for using  induction hypotheses and recursive function definitions for guiding induction. 
We present  exhaustive experimental results demonstrating the practical impact of our approach as implemented within Vampire.
\ifbool{shortversion}{}{

This is an extended version of a  Principles of Systems Design 2022 paper with the same title and the same authors.}%
\keywords{Induction \and Formal Verification \and Theorem Proving}
\end{abstract}

\section{Introduction}\label{sec:intro}

One commonly used theory in the development of imperative/functional programs is the theory of inductively defined
data types, such as natural numbers (e.g. see Figure~\ref{fig:ex1figure}(a)). Automating reasoning in formal verification  therefore also needs to automate induction. 
Previous works on
automating induction mainly focus on inductive theorem
proving~\cite{ACL2,Rippling93,HipSpec,Zeno,Imandra}: deciding when induction should be applied and what induction axiom should be
used.  
Recent advances related to automating inductive reasoning, such as first-order reasoning with inductively defined data types~\cite{POPL17Kovacs},  inductive
strengthening~\cite{ReynoldsK15} and  structural 
induction in
superposition~\cite{Peltier13,Cruanes17,CADE19,CICM20,Peltier20},
open up new possibilities for 
automating 
induction. In this paper we survey our recent results  towards   automating inductive reasoning for  first-order properties with inductively defined data types and beyond.

\noindent{\textbf{Relation to the state-of-the-art.}} Our work 
automates induction by integrating it directly in the saturation-based  
approach of first-order  provers~\cite{CAV13,E19,Spass09}. These provers implement saturation-based proof search using the
superposition
calculus~\cite{NieuwenhuisRubio:HandbookAR:paramodulation:2001}. Moreover, they 
rely on powerful indexing algorithms, notions of redundancy,
selection functions and term orderings for making  theorem proving efficient. First-order theorem provers complement SMT solvers in reasoning with theories and quantifiers, as evidenced in the annual
system competitions of SMT solvers~\cite{SMTComp,SMTComp_Survey} and first-order provers~\cite{casc}.

 Our approach towards automating induction is conceptually different from previous
attempts to use induction with superposition
\cite{Peltier13,Cruanes17,Peltier20}, as we are not restricted to specific
clause 
splitting algorithms and heuristics~\cite{Cruanes17}, nor are we limited to induction over inductively defined data types using a subterm ordering~\cite{Peltier20}. As a
result, we stay within the standard saturation framework and do not
have to introduce constraint clauses, additional predicates, nor change
the notion of redundancy as in~\cite{Peltier20}. In addition, our approach can be used to automate induction over arbitrary, and not just inductively defined, data types, such as integers (Section~\ref{sec:integers}). 
Our work is  also fundamentally different from rewrite-based approaches automating
induction~\cite{ACL2,Rippling93,Imandra,HipSpec,ReynoldsK15,Zeno}, as we do
not rely on external algorithms/heuristics to generate subgoals/lemmas
of an inductive property. Instead, applications of induction become
inference rules of the saturation process, adding instances of
appropriate induction schemata. 
We extend superposition
reasoning with new inference rules capturing inductive steps (Sections~\ref{sec:induction_rules}-\ref{section:indmc}),  
and optimize the saturation theorem proving process with induction. In addition, we instantiate induction axioms with logically stronger versions of
the property being proved and use induction hypotheses as specialized rewrite rules (Section~\ref{sec:extensions}). 

This combination of saturation with induction is very powerful.  Our experimental
results show that many problems previously unsolved by any system can be
 solved by our work, some resulting in very complex proofs
of program properties and proofs of complex mathematical properties (Section~\ref{sec:experiments}).
\ifbool{shortversion}{}{ 
Some of these proofs involve, among
tens of thousands of superposition inferences, over 100 applications of
induction.
} 
\vskip1em

\noindent{\textbf{Contributions.}} This paper serves as  \emph{a survey} of our recent progress in automating induction using a first-order theorem prover~\cite{CADE19,CICM20,CADE21,FMCAD21}.

\begin{itemize} 
\item We give a small tutorial of induction in saturation, 
helping  non-experts in theorem proving to understand and further use our methodology. To this end, we describe saturation theorem proving and the main concepts of saturation with induction (Sections~\ref{sec:saturation}-\ref{sec:induction_rules}). 

\item
We overview technical considerations for turning saturation with induction into an efficient approach (Section~\ref{sec:induction_rules})\ifbool{shortversion}{ and discuss variants of induction inference rules (Section~\ref{sec:induction:TA}).}{.  We  discuss variants of induction inference rules
over inductively defined data types (Section~\ref{sec:induction:TA}) and integers (Section~\ref{sec:integers}). }

\item We present  extensions of induction inference rules with multiple premises (Section~\ref{section:indmc}),   generalizations and integer reasoning (Section~\ref{sec:extensions}).

\item We report on exhaustive experiments comparing and analysing our approach to state-of-the-art methods (Section~\ref{sec:experiments}). 
\end{itemize}

\section{Motivating Example}\label{sec:motivating}
We motivate the challenges of automating induction for formal verification using the functional program of Figure~\ref{fig:ex1figure}(a). 
This program defines the inductively defined data type {\tt nat} of natural numbers. In first-order logic, 
 this data type corresponds to a term
algebra with constructors {\tt 0} (zero) and {\tt s} (successor);  inductively defined data types, such as {\tt nat}, are special cases of
term algebras.
The functional program in 
Figure~\ref{fig:ex1figure}(a)  implements  {\tt add},  {\tt even} and  {\tt half} operations  over naturals, 
by using recursive equations (function definitions) preceded by the {\bf fun} construct. These recursive equations correspond to universally quantified equalities in first-order logic, as listed in the axioms of Figure~\ref{fig:ex1figure}(b). 

\begin{figure*}[t]
  \begin{minipage}{.405\textwidth}
  \reserved{assume}~ $\even(x)$\\[.5em]
  \textbf{datatype}~ $\nat = \zero \mid \suc(x)$
  \begin{flalign*}
    \textbf{fun } &\add (\zero, y) = y &\\
      |\ &\add(\suc(z), y) = \suc(\add(z,y)); &\\
    \textbf{fun } &\even(\zero) = \top &\\
     |\ &\even(\suc(z)) = \neg\even(z); &\\
    \textbf{fun } &\half(\zero) = \zero &\\
     |\ &\half(\suc(\zero)) = \zero &\\
     |\ &\half(\suc(\suc(z))) = \suc(\half(z)); &
  \end{flalign*}
  \reserved{assert}~ $x=\add(\half(x),\half(x))$\\
  {\phantom{wwhite-space}} (a)
  \end{minipage}%
  \vline\hfill%
  \begin{minipage}{0.005\linewidth}\,\end{minipage}%
  \begin{minipage}{0.58\linewidth}
    \vspace*{-1em}
    \begin{flalign*}
        &\text{Axiomatization of $\add$, $\even$ and $\half$:}&\\
        &\;\;\forall y \in \nat. (\add(\zero, y) = y) &\\
        &\;\;\forall z,y \in \nat. (\add(\suc(z), y) = \suc(\add(z,y))) &\\
        &\;\;\even(\zero) &\\
        &\;\;\forall z \in \nat. (\even(\suc(z)) \leftrightarrow \neg\even(z)) &\\
        &\;\;\half(\zero) = \zero &\\
        &\;\;\half(\suc(\zero)) = \zero &\\
        &\;\;\forall z \in \nat. (\half(\suc(\suc(z))) = \suc(\half(z))) &\\
        &\text{Verification task (conjecture): }&\\
        &\;\;\forall x\in\nat.(\even(x) \rightarrow 
        x=\add(\half(x),\half(x)))&
    \end{flalign*}%
   {\phantom{wwhite-space}} (b)
  \end{minipage}
\caption{Motivating example over inductively defined data
  types\label{fig:ex1figure}.}
\end{figure*}

The expected behaviour of Figure~\ref{fig:ex1figure}(a) is specified using program assertions in first-order logic: the pre-condition using the \reserved{assume} construct and the post-condition using \reserved{assert}. 
Figure~\ref{fig:ex1figure}(a) satisfies its requirements. 
Formally proving correctness of Figure~\ref{fig:ex1figure}(a) essentially requires proving the conjecture of Figure~\ref{fig:ex1figure}(b), 
establishing that $\half(x)$ of an {\tt even} natural number $x$ added to $\half(x)$ equals the original number $x$. That is, 
\begin{equation}
\forall x\in\nat. \big(\even(x)\rightarrow x=\add(\half(x),\half(x))\big).\label{eq:ex1}
\end{equation}

Proving~\eqref{eq:ex1}, and thus establishing correctness of Figure~\ref{fig:ex1figure}(a), is however challenging as
it requires   induction over the naturals. As such,  finding and using an appropriate induction schemata is needed.  
The following sound \textit{structural induction schema} for a formula $F$ could, for example, be used, where $F$ contains (multiple occurrences of) a natural-valued variable $x$:
\begin{equation}\label{eq:nat-schema1}\Big(F[\zero]\land \forall z\in\nat.(F[z]\rightarrow F[\suc(z)])\Big)\rightarrow \forall x\in\nat. F[x]\end{equation}

We instantiate schema~\eqref{eq:nat-schema1} by considering $\forall x\in\nat. F(x)$ to be formula~\eqref{eq:ex1}, yielding the induction formula:
\begin{equation}\label{eq:ex1.1}
\begin{array}{ll}
  \texttt{(IB)}\quad&  \big(\even(\zero)\rightarrow\zero=\add(\half(\zero),\half(\zero))\big)\land \\[.5em]
  \texttt{(IS)} & 
\forall z\in\nat.\begin{pmatrix}
\big(\even(z)\rightarrow z=\add(\half(z),\half(z))\big)\rightarrow\\ \big(\even(\suc(z))\rightarrow \suc(z)=\add(\half(\suc(z)),\half(\suc(z)))\big)
\end{pmatrix} \\[1em]
& \rightarrow\forall x\in\nat.\even(x)\rightarrow x=\add(\half(x),\half(x)),\end{array}\end{equation}
where the subformulas denoted by {\tt (IB)} and {\tt (IS)} correspond to the \emph{induction base case} and the \emph{induction step case} of~\eqref{eq:ex1.1}. 
Since schema~\eqref{eq:nat-schema1} is sound, its instance~\eqref{eq:ex1.1} is  valid. As such,  the task of proving~\eqref{eq:ex1} is reduced to  proving the base case and step case~of \eqref{eq:ex1.1}.

Using the definitions of $\half$ and $\add$ from Figure~\ref{fig:ex1figure}(b), the base case {\tt (IB)}  simplifies to the tautology $\top\rightarrow\zero = \zero$. 
On the other hand, proving {\tt (IS)} requires additional inductive reasoning. Yet, the induction scheme~\eqref{eq:nat-schema1} cannot be  used as  $\even(z)$ and $\even(\suc(z))$ yield two different base cases. 
We overcome this limitation by using an additional induction schema with two base cases, as follows: 
\begin{equation}\label{eq:nat-schema2}
\big(F[\zero]\land F[\suc(\zero)]\land \forall z.(F[z]\rightarrow F[\suc(\suc(z))])\big)\rightarrow \forall x. F[x]\end{equation}
As before, by instantiating~\eqref{eq:nat-schema2} with~\eqref{eq:ex1} and simplifying based on the axioms of Figure~\ref{fig:ex1figure}(b),  
we are left with proving the step case:
{\small
\begin{equation}\label{eq:ex1.2}
\begin{array}{ll}
\texttt{(IH)}
\quad\forall z\in&\nat.
\Big(\big(\even(z)\rightarrow z=\add(\half(z),\half(z))\big)\rightarrow\\ 
\texttt{(IC)}&\big(\even(\suc(\suc(z)))\rightarrow\suc(\suc(z))=\add(\half(\suc(\suc(z))),\half(\suc(\suc(z))))\big)\Big)\hspace{1em}
\end{array}
\end{equation}}%
The antecedent {\tt (IH)} and conclusion {\tt (IC)} of \eqref{eq:ex1.2} are called the \emph{induction (step) hypothesis} and \emph{induction step conclusion} of the step case, respectively. After rewriting $\even(\suc(\suc(z)))$ to $\even(z)$ in {\tt (IC)}, both {\tt (IH)} and {\tt (IC)} have the same assumption $\even(z)$, which can be discarded. By rewriting the remaining conclusions in {\tt (IH)} and {\tt (IC)} using  the definitions of $\half$ and $\add$, 
\ifbool{shortversion}{as well as the 
 the injectivity of the term algebra constructor~$\suc$, we obtain: 
}
{we obtain: 
\begin{equation}\label{eq:ex1.3}
\begin{array}{ll}
\texttt{(IH)}
\quad\qquad\forall z\in\nat.&\hspace{-0.2em}\big(z=\add(\half(z),\half(z))\rightarrow\\
\texttt{(IC)}&\qquad\;\;\suc(\suc(z))=\suc(\add(\half(z),\suc(\half(z))))\big)\hspace{2em}
\end{array}
\end{equation}
We simplify {\tt (IC)} in \eqref{eq:ex1.3} by the injectivity of the term algebra constructor~$\suc$:
}
\begin{equation}\label{eq:ex1.4}
\begin{array}{ll}
\texttt{(IH)}
\quad\qquad\forall z\in\nat.&\hspace{-0.2em}\big(z=\add(\half(z),\half(z))\rightarrow\\
\texttt{(IC)}&\qquad\suc(z)=\add(\half(z),\suc(\half(z)))\big)\hspace{5em}
\end{array}
\end{equation}
Since the more complex right-hand side of {\tt (IH)} is not equal to any subterm of {\tt (IC)} in \eqref{eq:ex1.4}, we have to use {\tt (IH)} in the left-to-right direction -- in order to preserve validity, our only option is to rewrite $z$ on the left-hand side of {\tt (IC)}:
\begin{equation}\label{eq:ex1.5}
    \forall z\in\nat.\big(\suc(\add(\half(z),\half(z)))=\add(\half(z),\suc(\half(z)))\big)
\end{equation}
Equation~\eqref{eq:ex1.5} is a special case of the  formula $\forall x,y\in\nat.\suc(\add(x,y)) = \add(x,\suc(y))$ which can be easily verified using the induction schema~\eqref{eq:nat-schema1}. This establishes the correctness of Figure~\ref{fig:ex1figure}(a). 

The verification task of Figure~\ref{fig:ex1figure}(a) highlights the main difficulties in automating inductive reasoning: (i) incorporating induction into saturation (Section~\ref{sec:induction_rules}); (ii) finding suitable induction schemata 
\ifbool{shortversion}{(Section~\ref{sec:induction:TA})}
{(Section~\ref{sec:induction:TA} and Section~\ref{sec:integers})}; and (iii)  using extensions of induction inference rules to further push the boundaries of automating induction (Sections~\ref{section:indmc}--\ref{sec:extensions}). We next  present our solutions to these challenges, based on our results from~\cite{CADE19,CICM20,CADE21,FMCAD21}. 

\section{Preliminaries}\label{sec:prelim}
We assume familiarity with \textit{standard multi-sorted first-order logic with equality}.
Functions are denoted with $f$, $g$, $h$, predicates with $p$, $q$, $r$, variables with
$x$, $y$, $z$, $w$, and Skolem constants with $\sigma$, all possibly with indices.
A term is \textit{ground} if it contains no variables. \ifbool{shortversion}{}{By $\overline{x}$ and $\overline{t}$ we denote tuples of variables and terms, respectively.}
We use the words \textit{sort} and \textit{type} interchangeably. We distinguish special sorts
called \textit{term algebra sorts}, function symbols for term algebra sorts called \textit{constructors}
and \textit{destructors}. 
For a term algebra sort $\tau$, we denote its constructors with $\Sigma_\tau$. For each $c\in \Sigma_\tau$, we denote its arity with $n_c$ and the corresponding destructor returning the value of the $i$th argument of $c$ by $d_c^i$. Moreover, we denote with $P_c$ the set of argument positions of $c$ of the sort $\tau$.
We say that 
$c$ is a \textit{recursive constructor} if $P_c$ is non-empty, otherwise it is called a \textit{base constructor}.
We call the ground terms built from
the constructor symbols of a sort its \textit{term algebra}. 
We axiomatise term algebras using their \textit{injectivity}, \textit{distinctness},
\textit{exhaustiveness} and \textit{acyclicity}
axioms~\cite{POPL17Kovacs}. 
We refer to term algebras
also as algebraic data types or inductively defined data types.
Additionally, we assume a distinguished
\emph{integer sort}, denoted by~$\intg$. When we use standard integer predicates
$<$, $\leq$, $>$, $\geq$, functions $+, -, \dots$ and constants $0, 1, \dots$, we
assume that they denote the corresponding interpreted integer predicates and functions with their
standard interpretations. All other symbols are uninterpreted.

We use the standard logical connectives $\neg$, $\lor$, $\land$, $\rightarrow$ and $\leftrightarrow$,
and quantifiers $\forall$ and $\exists$. We write
quantifiers like $\forall x\in\tau$ to denote that $x$ has the sort~$\tau$ where it is not clear from the context.
A \textit{literal} is an atom or its negation.
For a literal $L$, we write
$\overline{L}$ to denote its complementary literal. 
A disjunction of literals is a \textit{clause}. We denote clauses by $C, D$ and reserve the symbol $\square$ for the \textit{empty clause}
which is logically equivalent to $\bot$. We denote the \textit{clausal normal form} of a formula $F$
by $\mathtt{cnf}(F)$. We call every term, literal, clause or formula an \textit{expression}.
\ifbool{shortversion}{}{We use
the notation $s\trianglelefteq t$ to denote that $s$ is a \textit{subterm} of $t$
and $s\triangleleft t$ if $s$ is a \textit{proper subterm} of $t$.}

We write $E[s]$ to denote that the expression $E$ contains $k$ distinguished occurrence(s) of the term
$s$, with $k\geq 0$.
For simplicity, $E[t]$ means that these occurrences of $s$ are replaced by the term $t$.
A \textit{substitution} $\theta$ is a mapping from variables to terms. A substitution $\theta$ is a \textit{unifier} of two terms $s$ and $t$ if $s\theta= t\theta$, and is a \textit{most general unifier} (\textit{mgu}) if for every unifier $\eta$ of $s$ and $t$, there exists substitution $\mu$ s.t. $\eta=\theta\mu$. We denote the mgu of $s$ and $t$ with $\mathtt{mgu}(s,t)$.


\section{Saturation-Based Theorem Proving}\label{sec:saturation}
We briefly introduce saturation-based proof search,
which is the leading technology  for automated
first-order theorem proving. For  details, we refer to~\cite{CAV13}.

First-order theorem provers work with clauses, rather than with
arbitrary formulas. Given a set $S$ of input clauses, first-order
provers {\it saturate} $S$ by computing all logical consequences of $S$
 with respect to a sound inference
system $\mathcal{I}$. The saturated set of $S$ is called the {\it closure} of $S$
and the process of computing the closure of $S$ is called
\textit{saturation}. 
If the closure of $S$ contains the empty clause $\square$, the original
set $S$ of clauses is unsatisfiable.
A simplified saturation algorithm for a sound inference system $\mathcal{I}$
is given in Algorithm~\ref{alg:saturation}, with a clausified goal $B$ ($\neg B$ is also clausified) and clausified assumptions $A$ as input.


\begin{algorithm}[t]
\caption{The Saturation Loop.\label{alg:saturation}}
\begin{tabbing}
	{\scriptsize 1}\quad initial set of clauses $S:=A\cup\{\neg B\}$\\ 
	{\scriptsize 2}\quad \reserved{\texttt{repeat}}\\
	{\scriptsize 3}\quad\quad Select  clause $G\in S$\\
	{\scriptsize 4}\quad\quad Derive consequences ${C_1,\ldots,C_n}$ of $G$ and formulas from $S$ using rules of $\mathcal{I}$\\
	{\scriptsize 5}\quad\quad $S:=S\cup \{C_1,\ldots,C_n\}$\\
	{\scriptsize 6}\quad\quad \reserved{\texttt{if}} $\square\in S$ \reserved{\texttt{then}}~ \reserved{\texttt{return}}~ $A\implies B$ is UNSAT\\
	{\scriptsize 8}\quad \reserved{\texttt{return}}~ $A\implies B$ is SAT
\end{tabbing}
\end{algorithm}

Note that a saturation algorithm proves validity of $B$ by establishing unsatisfiabiliy of $\neg B$ using the assumptions $A$; we refer to this proving process as a {\it refutation} of $\lnot B$ from $A$. 
Completeness and efficiency of saturation-based reasoning rely
heavily on properties of selection and addition of clauses from/to $S$, using the inference system $\mathcal{I}$ (lines 3--5). 
To organize saturation,
first-order provers use simplification {\it orderings}
on terms, which are extended to
orderings over literals and clauses; for simplicity, we write
$\succ$ for both the term ordering and its clause/multiset ordering 
extensions. Given an ordering $\succ$, a clause $C$ is
\textit{redundant} with respect to a set $S$ of clauses
 if there exists a subset $S'$ of $S$ such that $S'$ is smaller
than $\{C\}$, that is $\{C\}\succ S'$  and $S'\implies C$.

The \textit{superposition calculus}, denoted as \Sup,  is the most common inference system employed
by saturation-based first-order theorem provers for first-order logic with equality~\cite{NieuwenhuisRubio:HandbookAR:paramodulation:2001}. 
A summary of superposition inference rules is given in Figure~\ref{fig:sup}. 
The superposition calculus \Sup{} is \textit{sound} and
\textit{refutationally complete}: for any unsatisfiable formula $\neg B$, the
empty clause can be derived as a logical consequence of $\neg B$. 

\ifbool{shortversion}{
}{ 
In addition to the rules of Figure~\ref{fig:sup}, modern saturation-based theorem provers \cite{E19,CAV13,Cruanes17} using the superposition calculus also implement special cases of superposition, with the aim of keeping the search space $S$ small.
To this end, the general theory of redundancy is exploited ensuring that redundant
clauses can be eliminated during proof search without destroying
completeness of the \Sup calculus.
}

\begin{figure}[tb]
\textbf{Superposition:}
\begin{align*}
	&\infer[]{(L[r]\lor C \lor D)\theta}{l=r\lor C\quad L[l^\prime]\lor D}&
	&\infer[]{(s[r]\neq t\lor C \lor D)\theta}{l=r\lor C\quad s[l^\prime]\neq t\lor D}&
	\infer[]{(s[r]= t\lor C \lor D)\theta}{l=r\lor C\quad s[l^\prime]= t\lor D}
\end{align*}
where $\theta:=\mathtt{mgu}(l,l^\prime)$, $r\theta\not\succeq l\theta$, (first rule only) $L[l^\prime]$ is not an equality literal, and (second and third rules only) $t\theta\not\succeq s[l^\prime]\theta$.
\vspace*{-1em} \\
\begin{minipage}{.3\textwidth}
\textbf{Binary resolution:}
\begin{equation*}
	\infer[]{(C \lor D)\theta}{L\lor C\quad\neg L^\prime\lor D}
\end{equation*}
where $\theta:=\mathtt{mgu}(L,L^\prime)$.\end{minipage}\begin{minipage}{.01\textwidth}
\phantom{a}
\end{minipage}\begin{minipage}{.3\textwidth}
\textbf{Equality resolution:}
\begin{equation*}
	\infer[]{C\theta}{s\not=t\lor C}
\end{equation*}
where $\theta:=\mathtt{mgu}(s,t)$.
\end{minipage}\begin{minipage}{.03\textwidth}
\phantom{a}
\end{minipage}\begin{minipage}{.3\textwidth}
\vspace{2.15em}
\textbf{Equality factoring:}
\begin{equation*}
	\infer[]{(s=t\lor t\not=t' \lor C)\theta}{s=t\lor s'=t' \lor C}
\end{equation*}
where $\theta:=\mathtt{mgu}(s,s')$, $t\theta\not\succeq s\theta$, and $t'\theta\not\succeq t\theta$.\\
\end{minipage}
\caption{The superposition calculus \Sup{} for first-order logic with equality\label{fig:sup}.}
\end{figure}

\section{Saturation with Induction}\label{sec:induction_rules}
We now describe our approach towards automating inductive reasoning within saturation-based proof search. We illustrate the key ingredients of our method using our motivating example from Figure~\ref{fig:ex1figure}(a), that is proving~\eqref{eq:ex1} in order to establish correctness of Figure~\ref{fig:ex1figure}(a). 
As mentioned in Section~\ref{sec:saturation}, proving~\eqref{eq:ex1} in a saturation-based approach means refuting the clausified negation of~\eqref{eq:ex1}, that is, refuting the following two clauses: 
\begin{align}
    &\even(\sigma_0)\label{eq:ex1.cl1}\\
    &\sigma_0 \neq\add(\half(\sigma_0),\half(\sigma_0))\label{eq:ex1.cl2}
\end{align}

We establish invalidity of inductive formulas, such as~\eqref{eq:ex1.cl1}-\eqref{eq:ex1.cl2}, by \emph{integrating the application of induction as additional inference rules of the saturation process}. Our {induction inference rules} are  used directly in Algorithm~\ref{alg:saturation}, as follows: 
\begin{itemize}
\item[(i)] we pick up an inductive property $G$ in the search space $S$ (line 3); 
\item[(ii)] derive new induction axioms $C_1, \ldots, C_n$ (instances of {\it induction schemata}), aiming at refuting $G$, or sometimes a more general formula than $G$ (line 4);  
\item[(iii)] add the induction axioms $C_1, \ldots, C_n$ to the search space (line 5). 
\end{itemize}
Our work therefore follows a different approach than the one used in inductive theorem provers, as we do not rely on 
external algorithms to generate subgoals/stronger formulas $G'$ of an  inductive property $G$ nor do we replace $G$ by subgoals/stronger formulas $G'$. Rather, new induction axioms $C_i$, and sometimes new induction axioms $C_i'$ for more general formulas $G'$, are derived from $G$ and used in the search space $S$ \emph{in addition} to $G$.

Finding the right induction schema and developing efficient induction inference rules for deriving inductive axioms/formulas (steps (i)-(ii) above) are  crucial for saturation with induction. In~\cite{CADE19} we introduced the 
following induction inference rule, parametrized by a valid induction schema:  
\begin{equation*}
\infer[\small{\tt (Ind)},]{\cnf(F \rightarrow \forall x.L[x])}{\overline{L}[t] \lor C}
\end{equation*}
where $t$ is a ground term, $L$ is a ground literal, $C$ is a clause, and $F \rightarrow \forall x.L[x]$
is a valid induction schema.
For example, the 
induction schema~\eqref{eq:nat-schema1} for $F$ can be used in {\tt (Ind)}. We call $\overline{L}[t]$ the \emph{induction literal} and $t$ the \emph{induction term}. 
We note that {\tt (Ind)} can  naturally be generalized to handle multiple induction terms, as in~\cite{FMCAD21}. 
In this paper, we only use the rule with one induction term.

Based on Algorithm~\ref{alg:saturation} (the
saturation-based proof search algorithm), note that the application of 
{\tt (Ind)} adds new clauses to the search space by clausifying induction formulas ({\tt cnf()} in {\tt (Ind)}). These new clauses 
then become potential candidates to be selected in the next steps of the 
algorithm. As such, the selection of these  new clauses are likely to be delayed, and thus their use in proving an inductive goal becomes highly inefficient. We therefore propose the application of {\tt (Ind)} followed by a binary resolution step  to ``guide" induction
over selected induction literals and terms. In particular, upon the application of  {\tt (Ind)}, we do not add $\cnf(F \rightarrow \forall x.L[x])$ to the search space. Instead, we binary resolve the conclusion literal $L[x]$ against $\overline{L}[t]$, allowing us to only add the  formula $\cnf(\lnot F)\lor C$ to the search space, whenever {\tt (Ind)} is applied.  

In order to ``guide" and combine the application of {\tt (Ind)} with a binary resolution rule,  we exploit instances of {\tt (Ind)} for special cases of induction schemata over term algebras (Section~\ref{sec:induction:TA}) and integers (Section~\ref{sec:integers}). We also consider extension of 
{\tt (Ind)} for more general and efficient inductive reasoning (Section~\ref{section:indmc}--\ref{sec:extensions}).

\section{Induction with Term Algebras} \label{sec:induction:TA}

We first consider the theory of term algebras and introduce   instances of the induction rule {\tt (Ind)}, by exploiting properties of the induction literal $\overline{L}[t]$ and induction schemata over the induction term $t$. For now,  the induction term $t$ is  a ground element from a term algebra. 

\paragraph{Structural Induction.}
The first instance of {\tt (Ind)} uses the following  constructor-based structural induction schema, where $L[x]$ is a literal containing (possibly multiple occurrences of) $x$ of a term algebra sort $\tau$: 
\begin{equation}\label{eq:struct-schema}
  \big(\bigwedge_{c \in \Sigma_\tau} \forall y_1,...,y_{n_c}. (\land_{i\in P_c} L[y_i] \implies L[c(y_1,...,y_{n_c})])\big)
  \implies \forall x\in\tau.L[x]
\end{equation}
%
Note that the structural induction schema~\eqref{eq:nat-schema1} over naturals is an instance of~\eqref{eq:struct-schema}. 

\begin{example}
By instantiating schema \eqref{eq:struct-schema} with the sole literal of clause \eqref{eq:ex1.cl2} and induction term $\sigma_0$, we obtain:
{\small
\begin{equation}
\begin{split}
\begin{pmatrix}
    \zero =\add(\half(\zero),\half(\zero))\;\land\\
    \forall z\in\nat.\begin{pmatrix}z =\add(\half(z),\half(z))\rightarrow\\\suc(z) =\add(\half(\suc(z)),\half(\suc(z)))\end{pmatrix}\hspace*{-0.25em}
\end{pmatrix} 
    \rightarrow
    \begin{matrix}
    \forall x\in\nat.\big(x=\hspace*{4.6em}\\\add(\half(x),\half(x))\big)
    \end{matrix}
\end{split}\label{eq:ex1.f1}
\end{equation}
}%
The clausified form of~\eqref{eq:ex1.f1} consists of the following two clauses:
{\small
\begin{align*}
    \zero \neq\add(\half(\zero),\half(\zero))&\lor\sigma_1 =\add(\half(\sigma_1),\half(\sigma_1)) \lor x =\add(\half(x),\half(x))\\
    \zero \neq\add(\half(\zero),\half(\zero))&\lor\suc(\sigma_1)\neq\add(\half(\suc(\sigma_1)),\half(\suc(\sigma_1))) \\
    &\hspace*{13em}\lor x =\add(\half(x),\half(x))
\end{align*}}%
After applying {\tt (Ind)} instantiated with~\eqref{eq:ex1.f1} on~\eqref{eq:ex1.cl2}, the 
above clauses are resolved with the literal in clause~\eqref{eq:ex1.cl2}, adding to the search space the resulting clauses:
{\belowdisplayskip=-1em
\begin{align*}
    &\zero \neq\add(\half(\zero),\half(\zero))\lor\sigma_1 =\add(\half(\sigma_1),\half(\sigma_1))\\
    &\zero \neq\add(\half(\zero),\half(\zero))\lor\suc(\sigma_1)\neq\add(\half(\suc(\sigma_1)),\half(\suc(\sigma_1)))
\end{align*}\qed
}
\end{example}


\paragraph{Well-Founded Induction.} Two other instances of {\tt (Ind)}  exploit well-founded induction schemata, by using a binary well-founded relation $R$ on a term algebra $\tau$. For such an $R$, if there does not exists a smallest value $v\in\tau$ w.r.t. $R$ such that $L[v]$ does not hold, then $L[x]$ holds for any $x\in\tau$. This principle is formalized by the following schema:
\begin{equation}\label{eq:wf-schemaR}
    \Big(\neg\exists y\in\tau.\big(\neg L[y]\land \forall z\in\tau.(R(y,z) \rightarrow L[z])\big)\Big)\rightarrow\forall x\in\tau.L[x]
\end{equation}
However, to instantiate~\eqref{eq:wf-schemaR}, we need to find an $R$ suitable for the considered $\tau$.

Similarly to~\cite{CVC4induction}, we first consider the direct subterm relation expressed using term algebra constructors and destructors of the term algebra sort $\tau$.
We obtain the following instance of~\eqref{eq:wf-schemaR} to be applied in {\tt (Ind)}:
\begin{align}\label{eq:wf-schema1}
\begin{split}
    \Big(\neg\exists y.\big(\neg L[y]\land \hspace{-.5em}\bigwedge_{c\in \Sigma_\tau}(y=c(d_c^1(y),\dots,d_c^{n_c}(y))\rightarrow\bigwedge_{i \in P_c} &L[d_c^i(y)])\big)\Big)\rightarrow\forall x.L[x]
\end{split}
\end{align}
In the case of natural numbers, where $\pre$ is the destructor for $\suc$, we have the following instance of~\eqref{eq:wf-schema1} to be used in ({\tt Ind}): 
\begin{equation}\label{eq:nat-schema4}
    \Big(\neg\exists y\in\nat.\big(\neg L[y]\land (y=\suc(\pre(y)) \rightarrow L[\pre(y)])\big)\Big)\rightarrow\forall x\in\nat.L[x]
\end{equation}

Another instance of~\eqref{eq:wf-schemaR} to be used in {\tt (Ind)} 
employs a fresh predicate $\mathtt{less}_y$, as given next. The axiomatisation of such a predicate  enables  efficient reasoning over subterm properties withing saturation, as advocated  in~\cite{POPL17Kovacs}. 
\begin{align}\label{eq:nat-schema5}
\begin{split}
    \Big(\neg\exists y.\big(&\neg F[y]\land \forall z.(\mathtt{less}_y(z) \rightarrow F[z])\land (y=\suc(\pre(y))\implies\mathtt{less}_y(\pre(y)))\\[-.5em]
    &\land \forall w.(\mathtt{less}_y(\suc(\pre(w))) \implies \mathtt{less}_y(\pre(w)))\big)\Big)
    \rightarrow\forall x.F[x]
\end{split}
\end{align}




\paragraph{Induction with Recursive Function Definitions.}
In formalizing the induction schemata instances given e.g. in~\eqref{eq:nat-schema1} and~\eqref{eq:nat-schema4}, 
we considered the term algebra $\nat$ as an instance of $\tau$. 
To come up with the ``right" term algebra instance of $\tau$, we can 
also use terminating recursive function definitions from the input problem to be proven, such as  $\add$, $\even$ and $\half$ from Figure~\ref{fig:ex1figure}(a). The termination of such recursive functions naturally depends on a well-founded relation $R$. 

\ifbool{shortversion}{
}{ 
For an $n$-ary function $\mathtt{f}$ and a clausified function definition axiom $\mathtt{f}(\overline{s})=t\lor C$ in the search space, we call $\mathtt{f}(\overline{s})$ a \textit{function header} and any $\mathtt{f}(\overline{s^\prime})\trianglelefteq t$ a \textit{recursive call} of this function header. Moreover, we call an argument position $1\le i\le n$ \textit{inductive} if for any such function header-recursive call pairs, $s_i$ is a term algebra term (i.e. it only contains constructors and variables) and $s^\prime_i\triangleleft s_i$; in this case, $s_i$  is called an \emph{inductive argument}. Using inductive argument positions from  function definitions, we can then generate inductive schemata similar to \eqref{eq:struct-schema}, possibly with multiple induction terms.
}
\begin{example}\label{ex:recdef}
\ifbool{shortversion}{
We can obtain schema~\eqref{eq:nat-schema2} from $\half$ in Figure~\eqref{fig:ex1figure}(b) if we consider the well-founded relation based on its first argument. In particular, the third branch of $\half$ relates its first argument $\suc(\suc(z))$ to $z$ in its recursive call for all $z\in\nat$. This relation gives the step case of schema~\eqref{eq:nat-schema2}, and the base cases can be obtained by considering the terms in the first argument positions for the other two branches of $\half$.
}{
We can obtain schema~\eqref{eq:nat-schema1} from the second axiom of $\add$ in Figure~\eqref{fig:ex1figure}(b) with function header $\add(\suc(x),y)$ and recursive call $\add(x,y)$ due to $\suc(x)$ being a term algebra term and $x\triangleleft\suc(x)$. Moreover, the first argument of $\add$ in its first axiom gives the base case $\zero$.

Similarly, the induction step case of schema~\eqref{eq:nat-schema2} is given by the third axiom of $\half$ in Figure~\eqref{fig:ex1figure}(b) where the only argument $\suc(\suc(x))$ of the function header is a term algebra term and for the first argument of the recursive call $\half(x)$, we have $x\triangleleft\suc(\suc(x))$. Finally, the base cases  of schema~\eqref{eq:nat-schema2} are the first arguments of the function headers from the first two axioms of $\half$.
}

\ifbool{shortversion}{
Thus, based on the term $\half(\sigma_0)$ in clause \eqref{eq:ex1.cl2}, we can instantiate \eqref{eq:nat-schema2} inducting on term $\sigma_0$.
However, this induction axiom does}{
Thus, based on the function definitions in clauses \eqref{eq:ex1.cl1} and \eqref{eq:ex1.cl2}, we can instantiate both \eqref{eq:nat-schema1} and \eqref{eq:nat-schema2} inducting on term $\sigma_0$.
However, such induction axioms do} not yet lead to a refutation of~\eqref{eq:ex1}, because for each clausified induction axiom, new Skolem constants are introduced.
Thus, the literals in clauses resulting from applying {\tt (Ind)} on~\eqref{eq:ex1.cl1} or~\eqref{eq:ex1.cl2}, respectively, do not contain $\sigma_0$, and hence we cannot use~\eqref{eq:ex1.cl2} nor~\eqref{eq:ex1.cl1}, respectively, to refute them.
In the next section we therefore generalize {\tt (Ind)} towards the use of induction schemata with  multiple clauses.
\qed

\end{example}

\section{Multi-Clause Induction}\label{section:indmc}

Inducting on a single literal is sometimes not sufficient to get a refutation, as illustrated in Example~\ref{ex:recdef} for Figure~\ref{fig:ex1figure}(a). 
%
In general however, induction can be applied on literals from multiple clauses, 
similarly to formula~\eqref{eq:ex1.1} in Section~\ref{sec:motivating}. We 
generalize the inference rule {\tt (Ind)} towards multi-clause induction {\tt (IndMC)}: 
\begin{equation*}
\infer[\small{\tt (IndMC)}]{\cnf(F \rightarrow \forall x.(\bigwedge_{1\le i\le n}L_i[x]\rightarrow L[x]))}{L_1[t]\lor C_1\quad...\quad L_n[t]\lor C_n\quad\overline{L}[t] \lor C}
\end{equation*}
where $F \rightarrow \forall x.(\bigwedge_{1\le i\le n}L_i[x]\rightarrow L[x])$
is a valid induction formula, $\overline{L}$ and $L_i$ are ground literals and
$C$ and $C_i$ are clauses.  
Similarly to $\mathtt{(Ind)}$, our  new rule {\tt (IndMC)} is  used within saturation-based proof as an additional inference rule, followed by an application of binary resolution for guiding inductive reasoning. 

\begin{example}
We use schema~\eqref{eq:nat-schema2} with formula~\eqref{eq:ex1} with induction term $\sigma_0$ to instantiate {\tt (IndMC)} for premises \eqref{eq:ex1.cl1} and \eqref{eq:ex1.cl2}. The induction formula is:
{\footnotesize
\reqnomode
\begin{equation}\begin{split}\begin{pmatrix}\big(\even(\zero)\rightarrow \zero=\add(\half(\zero),\half(\zero))\big)\land\\
\big(\even(\suc(\zero))\rightarrow \suc(\zero)=\add(\half(\suc(\zero)),\half(\suc(\zero)))\big)\land\\
\forall z\hspace{-2pt}\in\hspace{-2pt}\nat.\hspace{-2pt}\begin{pmatrix}
\big(\even(z)\rightarrow z=\add(\half(z),\half(z))\big)\rightarrow\\
\big(\even(\suc(\suc(z)))\rightarrow \suc(\suc(z))=\add(\half(\suc(\suc(z))),\half(\suc(\suc(z))))\big)
\end{pmatrix}\end{pmatrix}\\
\rightarrow\forall x\in\nat. \big(\even(x)\rightarrow x=\add(\half(x),\half(x))\big)\end{split}\label{eq:ex1.f2}\end{equation}
}%
Clausification of formula~\eqref{eq:ex1.f2} results in twelve clauses, each containing the literals $\neg\even(x)$ and $x=\add(\half(x),\half(x))$, which we can binary resolve with clauses \eqref{eq:ex1.cl1} and \eqref{eq:ex1.cl2}. After simplifications are applied to the clauses from formula \eqref{eq:ex1.f2}, we are left with the following two clauses:
\begin{align}
    &\sigma_2=\add(\half(\sigma_2),\half(\sigma_2))\label{eq:ex1.cl7}\\
    &\suc(\sigma_2)\neq\add(\half(\sigma_2),\suc(\half(\sigma_2)))\label{eq:ex1.cl8}
\end{align}
We now need to rewrite \eqref{eq:ex1.cl8} with the induction hypothesis clause \eqref{eq:ex1.cl7} in the left-to-right orientation.
However, $\sigma_2\prec\add(\half(\sigma_2),\half(\sigma_2))$, which holds for any simplification ordering $\prec$, contradicts the superposition ordering conditions.
Moreover, even if we rewrote against the ordering, we would be left with
\begin{equation}
    \suc(\add(\half(\sigma_2),\half(\sigma_2)))\neq\add(\half(\sigma_2),\suc(\half(\sigma_2))),\label{eq:ex1.cl9.1}
\end{equation}
which is hard to refute using induction due to the induction term $\sigma_2$ occurring in the second argument of $\add$, which does not change in the recursive definition of $\add$ (see Figure~\ref{fig:ex1figure}(b)).
We overcome this limitation by extensions of inductive reasoning in Section~\ref{sec:extensions}.
\qed
\end{example}

\section{Extensions of  Inductions in Saturation}\label{sec:extensions}
\paragraph{Induction with Generalizations.}
It is common in mathematics that for proving a formula $A$, we prove instead a formula $B$ such that $B\implies A$. In other words, we prove a \emph{generalization} $B$ of $A$. 
Inductive theorem provers  implement various heuristics to  \textit{guess formulas/lemmas} $B$ and use $B$ instead of $A$  during proof search, see e.g.~\cite{Rippling93,HipSpec,ACL2,Imandra}. However, a saturation-based theorem prover
would not/can not do this, since goals/conjectures are not replaced by sub-goals in saturation-based proof search. We 
thus propose a different approach for implementing the common generalization recipe of mathematical theorem proving. 
Namely,  we  introduce the inference rule {\tt (IndGen)} of \emph{induction with generalization}, 
allowing us to (i)  add instances of induction schemata  not only for $A$ but also for versions of $B$ and then (ii) perform saturation over these induction schemata instances, using superposition reasoning.  Our {\tt (IndGen)} rule inducts only on  \textit{some} occurrences of the induction term $t$, as follows:

%

\begin{equation*}
\infer[\small{\tt (IndGen)},]{\cnf(F \rightarrow \forall x.L^\prime[x])}{\overline{L}[t] \lor C}
\end{equation*}
where $t$ is a ground term, $L$ is a ground literal, $C$ is a clause, $F \rightarrow \forall x.L^\prime[x]$
is a valid induction schema and $L^\prime[x]$ is obtained from $L[t]$ by replacing some occurrences of $t$ with $x$.

\begin{example}
We illustrate induction with generalization on the unit clause~\eqref{eq:ex1.cl9.1}.
One generalization that would help refute~\eqref{eq:ex1.cl9.1} by eliminating $\half(\sigma_2)$ is:
\begin{align}
    \forall x,y\in\nat.\suc(\add(x,y))=\add(x,\suc(y))\label{eq:ex1.gen2}
\end{align}
Instantiating schema~\eqref{eq:nat-schema1} with \eqref{eq:ex1.gen2} and variable $x$ would lead to a refutation when used with rule {\tt (IndGen)} on \eqref{eq:ex1.cl9.1}.
However, since we do not use $y$ from the generalization in the induction, there is no need to replace the occurrences of $\half(\sigma_2)$ corresponding to it in the generalized literal. Our final generalized induction formula, also leading to the refutation of~\eqref{eq:ex1.cl9.1}, is:
{\belowdisplayskip=-1em
\begin{align}\label{eq:ex1.f4}
\begin{split}
\begin{pmatrix}\suc(\add(\zero,\half(\sigma_2)))=\add(\zero,\suc(\half(\sigma_2)))\land\\
\forall z\in\nat.\begin{pmatrix}
\suc(\add(z,\half(\sigma_2)))=\add(z,\suc(\half(\sigma_2)))\rightarrow\\
\suc(\add(\suc(z),\half(\sigma_2)))=\add(\suc(z),\suc(\half(\sigma_2)))
\end{pmatrix}\end{pmatrix}\\\rightarrow\forall x\in\nat.\suc(\add(x,\half(\sigma_2)))=\add(x,\suc(\half(\sigma_2)))\end{split}
\end{align}
}
\qed
\end{example}

\paragraph{Rewriting with Induction Hypotheses.}
For turning saturation-based proof search into an efficient process, one key ingredient is to ensure that bigger terms/literals are rewritten by small ones (big/small w.r.t. the simplification ordering $\succ$), and not vice versa.
\ifbool{shortversion}{
However, this often prohibits using induction hypotheses to rewrite their corresponding conclusions which would be the necessary step to proceed with the proof.
}{ 
However, this often prohibits using induction hypotheses to rewrite their corresponding conclusions which would be the necessary step to proceed with the proof,
such as rewriting of~\eqref{eq:ex1.cl8} with~\eqref{eq:ex1.cl7} in the left-to-right orientation to obtain \eqref{eq:ex1.cl9.1}, on which we would then use {\tt (IndGen)} with induction formula \eqref{eq:ex1.f4} to proceed with the proof.
} 
To overcome this obstacle,  
we introduce the following  inference rule which uses an induction hypothesis literal to rewrite its conclusion: 
\begin{equation*}
\infer[\small{\tt (IndHRW)}]{\mathtt{cnf}(F \rightarrow \forall x.(s[r]=t)[x])}{l=r \lor D& s[l]\neq t \lor C}
\end{equation*}
where $s[l]\neq t$ is an induction conclusion literal with corresponding
induction hypothesis literal $l=r$,
$l\not\succeq r$, and $F \rightarrow \forall
x.(s[r]=t)[x]$ is a valid induction formula. 
Moreover, we resolve the clauses with the intermediate clause $s[r]\neq t\lor C\lor D$, obtained from the rewriting of the premises of $\mathtt{(IndHRW)}$. 

\begin{example} 
\ifbool{shortversion}{
}{} 
Using unit clause \eqref{eq:ex1.cl7} in a left-to-right orientation and rewriting the sides of unit clause \eqref{eq:ex1.cl8} one after the other, we get intermediate clauses, which are then used for generating induction formulas. One such intermediate clause is \eqref{eq:ex1.cl9.1}, from which the induction formula \eqref{eq:ex1.f4} is generated.
\ifbool{shortversion}{
After clausifying~\eqref{eq:ex1.f4}, a subsequent binary resolution is performed with intermediate clause~\eqref{eq:ex1.cl9.1}. By more simplifications using the definition of $\add$ and the injectivity of $\suc$, we finally obtain a refutation of~\eqref{eq:ex1}, concluding  thus  the correctness of  Figure~\ref{fig:ex1figure}(a).
}{
Below, the clausification of formula \eqref{eq:ex1.f4} is shown:
{\footnotesize
\begin{align}
    &\nonumber\suc(\add(\zero,\half(\sigma_2)))\neq\add(\zero,\suc(\half(\sigma_2)))\\
    \label{eq:ex1.f3.cl1}
    &\;\;\lor\suc(\add(\sigma_3,\half(\sigma_2)))=\add(\sigma_3,\suc(\half(\sigma_2)))\\
    &\nonumber\;\;\lor\suc(\add(x,\half(\sigma_2)))=\add(x,\suc(\half(\sigma_2)))\\[.5em]
    &\nonumber\suc(\add(\zero,\half(\sigma_2)))\neq\add(\zero,\suc(\half(\sigma_2)))\\
    \label{eq:ex1.f3.cl2}
    &\;\;\lor\suc(\add(\suc(\sigma_3),\half(\sigma_2)))\neq\add(\suc(\sigma_3),\suc(\half(\sigma_2)))\\
    &\nonumber\;\;\lor\suc(\add(x,\half(\sigma_2)))=\add(x,\suc(\half(\sigma_2)))
\end{align}
}%
By resolving both~\eqref{eq:ex1.f3.cl1} and~\eqref{eq:ex1.f3.cl2} with the intermediate clause~\eqref{eq:ex1.cl9.1}, we get the following clauses:
{\footnotesize
\begin{align}
    &\nonumber\suc(\add(\zero,\half(\sigma_2)))\neq\add(\zero,\suc(\half(\sigma_2)))\\
    \label{eq:ex1.f3.cl1.1}
    &\;\;\lor\suc(\add(\sigma_3,\half(\sigma_2)))=\add(\sigma_3,\suc(\half(\sigma_2)))\\[.5em]
    &\nonumber\suc(\add(\zero,\half(\sigma_2)))\neq\add(\zero,\suc(\half(\sigma_2)))\\
    \label{eq:ex1.f3.cl2.1}
    &\;\;\lor\suc(\add(\suc(\sigma_3),\half(\sigma_2)))\neq\add(\suc(\sigma_3),\suc(\half(\sigma_2)))
\end{align}
}%
After we rewrite the first literal, $\suc(\add(\zero,\half(\sigma_2)))\neq\add(\zero,\suc(\half(\sigma_2)))$, by the first axiom of $\add$,
we obtain the literal $\suc(\half(\sigma_2))\neq\suc(\half(\sigma_2))$ in both clauses, which we can remove as it is a trivially invalid inequality.
We are thus left only with the second literal from both~\eqref{eq:ex1.f3.cl1.1} and~\eqref{eq:ex1.f3.cl2.1}:
{\footnotesize
\begin{align}
    \label{eq:ex1.cl13}
    \suc(\add(\sigma_3,\half(\sigma_2)))&=\add(\sigma_3,\suc(\half(\sigma_2)))\\
    \label{eq:ex1.cl14}
    \suc(\add(\suc(\sigma_3),\half(\sigma_2)))&\neq\add(\suc(\sigma_3),\suc(\half(\sigma_2)))
\end{align}}%
We rewrite~\eqref{eq:ex1.cl14} by the second axiom of $\add$ twice, obtaining:
{\footnotesize
\begin{equation}
    \suc(\suc(\add(\sigma_3,\half(\sigma_2))))\neq\suc(\add(\sigma_3,\suc(\half(\sigma_2))))
\end{equation}}%
Using injectivity of $\suc$ we derive $\suc(\add(\sigma_3,\half(\sigma_2)))\neq\add(\sigma_3,\suc(\half(\sigma_2)))$, which we can finally resolve with~\eqref{eq:ex1.cl13}, resulting into $\square$.
For the whole formal proof of the assertion of Figure~\ref{fig:ex1figure}(a), we refer the reader to Appendix~\ref{sec:appendix1}.
}
\qed
\end{example}

\ifbool{shortversion}{
\paragraph{Integer Induction.}\label{sec:integers}
The last extension of our induction framework we introduce is \emph{integer induction}, motivated by the need of inductive reasoning in program analysis and verification problems using integers.
As the standard order $<$ (or $>$) over integers $\intg$ is not well-founded, we work with \emph{subsets of $\intg$ with a lower (and/or an upper) bound}.
We therefore define the \emph{downward, respectively upward, induction schema with symbolic bound $b$} as any formula of the form:
\begin{align*}
    \tag*{\textit{{\scriptsize (downward)}}}
  F[b]\land \forall y\in\intg. (y\leq b \land F[y] \implies F[y-1])
  \implies \forall x\in\intg.(x\leq b \implies F[x]); \\
    \tag*{\textit{{\scriptsize (upward)}}}
  F[b]\land \forall y\in\intg. (y\geq b \land F[y] \implies F[y+1])
    \implies \forall x\in\intg.(x\geq b \implies F[x]),
\end{align*}
respectively, where $F[x]$ is a formula with one or more occurrences of an integer variable $x$ and $b$ is an integer term not containing $x$ nor $y$.
Further, we also define \emph{interval downward, respectively upward, induction schema with symbolic bounds $b_1, b_2$} as any formula of the form:
{\small
\begin{equation*}
  \begin{array}{l}
  F[b_2]\land \forall y\in\intg. (b_1\hspace*{-0.1em}<\hspace*{-0.1em} y \hspace*{-0.1em}\leq\hspace*{-0.1em}  b_2 \land F[y] \implies F[y-1]) 
  \implies \forall x\in\intg.(b_1 \hspace*{-0.1em}\leq\hspace*{-0.1em} x \hspace*{-0.1em}\leq\hspace*{-0.1em} b_2 \implies F[x]); \; {\textit{{\scriptsize (down.)}}}\\
  F[b_1]\land \forall y\in\intg. (b_1 \hspace*{-0.1em}\leq\hspace*{-0.1em} y \hspace*{-0.1em}<\hspace*{-0.1em} b_2 \land F[y] \implies F[y+1])  
    \implies \forall x\in\intg.(b_1 \hspace*{-0.1em}\leq\hspace*{-0.1em} x \hspace*{-0.1em}\leq\hspace*{-0.1em} b_2\implies F[x]),\; {\textit{{\scriptsize (up.)}}}
  \end{array}
\end{equation*}}%
respectively, where $F[x]$ is a formula with one or more occurrences of an integer variable $x$ and $b_1,b_2$ are integer terms not containing $x$ nor $y$.%
\footnote{The above schemata can be seen as a special case of the multi-clause schemata used in the {\tt (IndMC)} rule from Section~\ref{section:indmc}, tailored specifically for integers.}

To automate inductive reasoning over integers, we need to automatically generate suitable instances of our integer induction schemata. To this end we introduce induction rules with the integer induction schemata in the conclusion, giving us the recipe for instantiating the schemata.
Since our schemata are sound, all resulting induction rules are sound as well.
When $t, b$ are ground terms and $L[t]$ is a ground literal,
the following is an \emph{integer upward induction rule}:
{\small
\begin{equation*}
  \infer[\small{(\texttt{IntInd}_\geq)}]{
    \cnf\Big(\big(L[b] \land \forall y\in\intg.(y\geq b \land L[y] \rightarrow L[y+1])\big) 
    \rightarrow \forall x\in\intg.(x\geq b\rightarrow L[x])\Big)
  }{
    \overline{L}[t] \lor C \; &\; t\geq b
  }
\end{equation*}}%
Our further integer induction rules using the other schemata are obtained similarly, as detailed in~\cite{CADE21}.

}{ 
\section{Integer Induction}\label{sec:integers}

\begin{figure}[t]
  \begin{minipage}{0.38\linewidth}
    \textbf{\underline{assume}~} $e\geq 0$
    \begin{flalign*}
        \textbf{fun } &\power(x, 0) = 1 &\\
         |\ &\power(x, e) = x \cdot \power(x, e-1); &
    \end{flalign*}
     \textbf{\underline{assert}~} $\power(2, e) > e$\\[1em]
       {\color{white}wwhite-space} (a)
  \end{minipage}%
  \vline\hfill%
  \begin{minipage}{0.02\linewidth}\,\end{minipage}%
  \begin{minipage}{0.58\linewidth}
    \vspace*{-1em}
    \begin{flalign*}
        &\text{Axiomatization of $\power$: }&\\
        &\;\;\forall x \in \intg. (\power(x, 0) = 1) &\\
        &\;\;\forall x, e \in\intg. (1\leq e \rightarrow \power(x, e) = x\cdot\power(x, e-1)) &\\
        &\text{Verification task (conjecture): }&\\
        &\;\;\forall e\in\intg.(0 \leq e \rightarrow e < \power(2, e))&
    \end{flalign*}
      {\color{white}wwhite-space} (b)
  \end{minipage}
  \caption{
    Functional program over integers.}
  \label{fig:power}
\end{figure}

In this section we introduce {\it integer induction} in saturation as a natural extension of our term algebra induction framework discussed so far. Inductive reasoning with integers is another common task in program analysis and verification, as  illustrated in Figure~\ref{fig:power}(a). 
%
The  first-order axiomatisation of the functional behavior and requirement for  Figure~\ref{fig:power}(a) is given in  Figure~\ref{fig:power}(b). 

The main insight of integer induction comes with the following observation of~\cite{CADE21}. As the standard order $<$ (or $>$) over integers $\intg$ is not not well-founded, we work with \emph{subsets of $\intg$ with a lower (and/or an upper) bound}.
We therefore define the \emph{downward, respectively upward, induction schema with symbolic bound $b$} as any formula of the form
\begin{align*}
    \tag*{\textit{{\scriptsize (downward)}}}
  F[b]\land \forall y\in\intg. (y\leq b \land F[y] \implies F[y-1])
  \implies \forall x\in\intg.(x\leq b \implies F[x]); \\
    \tag*{\textit{{\scriptsize (upward)}}}
  F[b]\land \forall y\in\intg. (y\geq b \land F[y] \implies F[y+1])
    \implies \forall x\in\intg.(x\geq b \implies F[x]),
\end{align*}
respectively, where $F[x]$ is a formula with one or more occurrences of an integer variable $x$ and $b$ is an integer term not containing $x$ nor $y$.
Further, we also define \emph{interval downward, respectively upward, induction schema with symbolic bounds $b_1, b_2$} as any formula of the form
\begin{align*}
    \tag*{\textit{{\scriptsize (downward)}}}
  \begin{split}
  F[b_2]\land \forall y\in\intg. (b_1 < y \leq b_2 \,\land \,&F[y] \implies F[y-1])  \\
    &\implies \forall x\in\intg.(b_1 \leq x \leq b_2 \implies F[x]);
  \end{split}\\
    \tag*{\textit{{\scriptsize (upward)}}}
  \begin{split}
  F[b_1]\land \forall y\in\intg. (b_1 \leq y < b_2 \,\land \,&F[y] \implies F[y+1])  \\
    &\implies \forall x\in\intg.(b_1 \leq x \leq b_2\implies F[x]),
  \end{split}
\end{align*}
respectively, where $F[x]$ is a formula with one or more occurrences of an integer variable $x$ and $b_1,b_2$ are integer terms not containing $x$ nor $y$.%
\footnote{The above schemata can be seen as a special case of the multi-clause schemata used in the {\tt (IndMC)} rule from Section~\ref{section:indmc}, tailored specifically for integers.}%

\begin{example}\label{ex:ind:intro}
Note that the verification task of Figure~\ref{fig:power}(b) holds also only over the non-negative subset of integers (lower bound 0). 
For proving correctness of Figure~\ref{fig:power}(a), we would therefore  use the following instance of the upward induction schema with symbolic bound with $b\bydef 0$ and $F[x]\bydef x<\power(2, x)$:
{\belowdisplayskip=-0.5em
\begin{equation}
  \begin{split}
    \big(0 < \power(2, 0)\land \forall x\in\intg. (x\geq 0 \land x < \power(2, x) \implies x+1 < \power(2, x+1))\big)\\
    \implies \forall y\in\intg.(y\geq 0 \implies y < \power(2, y)) \qquad\qquad\qquad\qquad
  \end{split}
\end{equation}}
\qed
\end{example}

To automate inductive reasoning over integers, we further need to automatically generate suitable instances of our integer induction schemata, for example upward induction schema instances for  Figure~\ref{fig:power}(a). Similarly as for term algebra reasoning, 
we  introduce induction rules with the integer induction schemata in the conclusion, which give us the recipe for instantiating the schemata.
Note that since our schemata above are sound, all our resulting induction rules are sound as well.
For brevity we only show the upward inference rules and leave out the symmetric downward rules.
When $t$ is a ground term, $b$ is a ground term and $L[t]$ a ground literal,
the following are \emph{integer upward induction rules}:
{\small
\begin{equation*}
  \infer[\small{(\texttt{IntInd}_\geq)}]{
    \cnf\Big(\big(L[b] \land \forall y\in\intg.(y\geq b \land L[y] \rightarrow L[y+1])\big) 
    \rightarrow \forall x\in\intg.(x\geq b\rightarrow L[x])\Big)
  }{
    \overline{L}[t] \lor C \; &\; t\geq b
  }
\end{equation*}
\begin{equation*}
  \infer[\small{(\texttt{IntInd}_{>})}]{
    \cnf\Big(\big(L[b] \land \forall y\in\intg.(y\geq b \land L[y] \rightarrow L[y+1])\big) 
    \rightarrow \forall x\in\intg.(x>b\rightarrow L[x])\Big)
  }{
    \overline{L}[t] \lor C \; &\; t> b
  }
\end{equation*}}%
While the $\texttt{IntInd}_{\geq}$ rule uses the upward schema exactly as defined above, the $\texttt{IntInd}_{>}$ rule uses a modified schema with weakened conclusion, containing $x>b$ instead of $x\geq b$.
This is a practical optimization: by resolving the clausified schema  against $\overline{L}[t]\lor C$, we obtain clauses containing $\lnot(t>b)$, which can be immediately resolved against the premise $t>b$.
If we instead needed to resolve away the literal $\lnot(t\geq b)$, we would need to first apply some theory reasoning to weaken $\lnot(t\geq b)$ into $\lnot(t>b)$.

Similar to the above rules with one bound, we introduce \emph{integer interval upward induction rules} for a ground term $t$, ground terms $b_1, b_2$ and a ground literal $L[t]$:
{\small
\begin{equation*}
  \infer[\small{(\texttt{IntInd}_{[\geq]})}]{
    \begin{array}{l}
    \mathtt{cnf}\Big(\big(L[b_1] \land \forall y\in\intg.(b_1 \leq y < b_2 \land L[y] \implies L[y+1])\big) \\
      \hspace*{3.5em} \implies \forall x\in\intg.(b_1 \leq x \leq b_2\implies L[x])\Big)
    \end{array}
  }{
    \overline{L}[t] \lor C \; &\; t\geq b_1 \; & \; t \leq b_2
  }
\end{equation*}
\begin{equation*}
  \infer[\small{(\texttt{IntInd}_{[>]})}]{
    \begin{array}{l}
    \mathtt{cnf}\Big(\big(L[b_1] \land \forall y\in\intg.(b_1 \leq y < b_2 \land L[y] \implies L[y+1])\big) \\
      \hspace*{3.5em} \implies \forall x\in\intg.(b_1 < x \leq b_2\implies L[x])\Big)
    \end{array}
  }{
    \overline{L}[t] \lor C \; &\; t> b_1 \; & \; t \leq b_2
  }
\end{equation*}}%
Note that in addition to the $\texttt{IntInd}_{[\geq]}$ and $\texttt{IntInd}_{[>]}$ rules, we can also introduce analogous rules $\texttt{IntInd}_{[\geq']}$ and $\texttt{IntInd}_{[>']}$ using the premise $t<b_2$ instead of $t\leq b_2$, and using correspondingly weakened conclusion.

Finally, we also introduce \emph{integer upward induction rule with default bound~0} for a ground term $t$ and a ground literal $L[t]$,
{\small
\begin{equation*}
  \infer[\small{(\texttt{IntInd}_{\geq0})},]{
    \cnf\Big(\hspace*{-0.18em}\big(L[0] \land \forall y\in\intg.(y\geq 0 \land L[y] \rightarrow L[y+1])\hspace*{-0.18em}\big) 
    \rightarrow \forall x\in\intg.(x\geq 0\rightarrow L[x])\Big)
  }{
    \overline{L}[t] \lor C
  }
\end{equation*}}%
which is together with the analogous integer downward induction rule with default bound 0 useful for proving properties holding for all integers.%
\footnote{See problem (13) in~\cite{CADE21}.}

\begin{figure}[t]
\begin{tabular}{l l r}
  ($C_1$)& $\power(x,0)=1$ & [input -- axiom] \\ 
  ($C_2$)& $e<1 \lor \power(x,e)=x\cdot \power(x, e-1)$ &  [input -- axiom] \\
  ($C_3$)& $0\leq \sigma_0$ & [input -- conjecture] \\
  ($C_4$)& $\power(2,\sigma_0)\leq\sigma_0$ & [input -- conjecture] \\
  ($C_5$)& $\power(2,0)\leq 0\lor 0\leq \sigma_1$  & [$\texttt{IntInd}_{\geq} C_4, C_3$, BR with $C_3, C_4$] \\
  ($C_6$)& $\power(2,0)\leq 0\lor \sigma_1<\power(2,\sigma_1)$  & [$\texttt{IntInd}_{\geq} C_4, C_3$, BR with $C_3, C_4$] \\
  ($C_7$)& $\power(2,0)\leq 0\lor \power(2,\sigma_1+1)\leq \sigma_1+1$ \qquad & [$\texttt{IntInd}_{\geq} C_4, C_3$, BR with $C_3, C_4$] \\
  ($C_8$)& $0\leq \sigma_1$  & [rewriting $C_5$ by $C_1$ and eval.] \\
  ($C_9$)& $\sigma_1<\power(2,\sigma_1)$  & [rewriting $C_6$ by $C_1$ and eval.] \\
  ($C_{10}$)& $\power(2,\sigma_1+1)\leq \sigma_1+1$  & [rewriting $C_7$ by $C_1$ and eval.] \\
  ($C_{11}$)& $1\leq \sigma_1\lor 0=\sigma_1$  & [$\leq$ axioms $C_8$] \\
  ($C_{12}$)& $1\leq \sigma_1\lor \power(2, 0+1)\leq 0+1$  & [rewriting $C_{10}$ by $C_{11}$] \\
  ($C_{13}$)& $1\leq \sigma_1\lor 2\cdot\power(2, 0)\leq 1\lor 1<1$  & [eval. $C_{12}$, rewriting by $C_2$, and eval.] \\
  ($C_{14}$)& $1\leq \sigma_1$  & [rewriting $C_{13}$ by $C_1$ and eval.] \\
  ($C_{15}$)& $2\cdot\power(2,\sigma_1)\leq \sigma_1+1$  & [rewriting $C_{10}$ by $C_2$ using $C_{14}$] \\
  ($C_{16}$)& $2\cdot\sigma_1< \sigma_1+1$  & [$<,\leq$ axioms $C_9,C_{15}$] \\
  ($C_{17}$)& $\sigma_1< 1$  & [cancellation of $\sigma_1$ in $C_{16}$] \\
  ($C_{18}$)& $\square$  & [BR $C_{14},C_{17}$]
%
\end{tabular}
  \caption{Key steps of a saturation-based proof certifying correctness of Figure~\ref{fig:power}(a)\label{tab:power:proof}.}
\end{figure}

\begin{example}\label{ex:integer:proof}
The key steps of proving the correctness of  Figure~\ref{fig:power}(a) using the induction rule $\texttt{IntInd}_{\geq}$ are displayed in Table~\ref{tab:power:proof}.
For clarity, we convert $\lnot(s<t)$ into $t\leq s$ and $\lnot(s\leq t)$ into $t<s$ for any terms $s, t$.
Clauses $C_1, C_2$ are the clausified axioms from Figure~\ref{fig:power}, while $C_3, C_4$ are the negated clausified conjecture from Figure~\ref{fig:power}.
We apply $\texttt{IntInd}_{\geq}$ on $C_4$ and $C_3$, producing clauses $C_5, C_6$ and $C_7$.
Next, we use superposition to rewrite the term $\power(2,0)$ in clauses $C_5, C_6, C_7$ by the definition from $C_1$ and then evaluate the resulting inequality $1\leq0$ to false, and we remove it, obtaining clauses $C_8, C_9, C_{10}$.
We then apply theory reasoning by applying binary resolution with suitable axiom for the predicate $\leq$, resulting into clause $C_{11}$.
Next we apply superposition on $C_{11},C_{10}$, obtaining $C_{12}$.
Then we evaluate the interpreted $0+1$ to $1$, apply superposition with $C_2$, and evaluate $1-1$ to $0$, obtaining $C_{13}$.
We next rewrite $C_{13}$ by $C_1$ and remove both $2\cdot1\leq 1$ and $1<1$ since they are evaluated to false, resulting into $C_{14}$.
We then rewrite $C_{10}$ by $C_2$ using superposition and binary resolution with $C_{14}$, obtaining $C_{15}$.
Using more theory reasoning, we arrive at $C_{17}$, which can be finally resolved against $C_{14}$, yielding the empty clause.\qed
%
\end{example}

} 

\section{Implementation and Experiments}\label{sec:experiments}
\subsection{Implementation}
\begin{table}[t]
\begin{tabular}{p{53mm}|p{67mm}}
  Name \& comma-separated values & Description \\ \hline\hline
  \texttt{--induction} \newline \phantom{}\hfill\texttt{int, struct, both, \underline{none}} & Enable induction on integers only, or induction on algebraic types only, or both, or none \\ \hline
\ifbool{shortversion}{
}{ 
  \texttt{--structural\_induction\_kind} \newline \phantom{}\hfill \texttt{\underline{one}, two, three, rec\_def, all} & What kind of induction axioms to use for induction on term algebras \\ \hline
  \texttt{--induction\_max\_depth} \newline \phantom{}\hfill\texttt{\underline{0}, 1, 2, \dots} & Maximum number of induction steps in any sequence of inferences, 0 means no maximum \\ \hline
  \texttt{--induction\_neg\_only} \hfill \texttt{\underline{on}, off} & Only apply induction on negative literals \\ \hline
  \texttt{--induction\_unit\_only} \hfill \texttt{\underline{on}, off} & Only apply induction on unit clauses \\ \hline
} 
  \texttt{--induction\_on\_complex\_terms} \newline \phantom{}\hfill \texttt{on, \underline{off}} & Apply induction also on complex terms \\ \hline
  \texttt{--induction\_multiclause} \hfill \texttt{\underline{on}, off} & Enable the \texttt{(IndMC)} form of induction rules \\ \hline
  \texttt{--induction\_gen} \hfill \texttt{on, \underline{off}} & Enable the \texttt{(IndGen)} form of induction rules \\ \hline
\ifbool{shortversion}{
}{ 
  \texttt{--induction\_hypothesis\_rewriting} \newline \phantom{}\hfill\texttt{\underline{on}, off} & Enable the \texttt{(IndHRW)} form of induction rules \\ \hline
  \texttt{--function\_definition\_rewriting} \newline \phantom{}\hfill\texttt{\underline{on}, off} & Use function definitions as rewrite rules with the intended orientation \\ \hline
  \texttt{--int\_induction\_interval} \newline \phantom{}\hfill\texttt{infinite, finite, \underline{both}} & Enable the integer induction rules, or interval integer induction rules, or both \\ \hline
  \texttt{--int\_induction\_default\_bound} \newline  \phantom{}\hfill\texttt{on, \underline{off}} & Enable the integer induction rules with default bound \\ \hline
} 
\end{tabular}
\vspace*{0.2em}
\caption{
\ifbool{shortversion}{
Selected induction options or \vampire{}. Default values are underlined.
}{ 
Summary of \vampire{}'s induction options. Default values are underlined.
} 
}
\label{tab:indoptions}
\end{table}

Our approach for automating induction in saturation is implemented in the  \vampire{} prover. 
All together, our implementation consists of around 7,800 lines of C++ code and is available online at \url{https://github.com/vprover/vampire/tree/int-induction}. 
In the following, \vampire{}* refers to the \vampire{} version supporting induction. 

Our induction rules allow us to derive many new clauses potentially leading to refutation of inductive properties. These new clauses
\ifbool{shortversion}{}{ -- especially in combination with theory reasoning in case of integer induction -- }%
might however pollute the search space without advancing the proof.
We therefore  introduce options to control the use of induction rules by inducting only on negative literals, unit clauses or clauses derived from the goal. 
Further, for induction over algebraic types, we only allow induction on terms containing a constant other than a base constructor.
\ifbool{shortversion}{
For integer induction, by default we disable rules with default bound, induction on interpreted constants, and induction on some comparison literals.
}{ 
For integer induction, by default we disable rules with default bound, and induction on interpreted constants.
Finally, by default we do not apply integer induction on $\overline{L}[t]\lor C$ if $\overline{L}[t]$ is in the form $t \circ s$ or $s \circ t$ where
$\circ \in \{<,\leq,>,\geq\}$ and $t$ does not occur in $s$.
} 
%
Our most relevant induction options are summarized in Table~\ref{tab:indoptions}.\footnote{\vampire{} also offers a so-called portfolio mode, in which it sequentially tries different option configurations for short amounts of time.} 

\subsection{Experimental Setup}
The main goal of our experiments was to evaluate how much  induction improves \vampire{}'s performance.
We therefore compared \vampire{}* to \vampire{} without induction.
We also show the numbers of problems solved by the SMT solvers \cvc4~\cite{CVC4induction}, \solver{Z3}~\cite{Z3}, where only \cvc4 supports induction. In our experiments, we do not include other provers, such as \acl2~\cite{ACL2} or \zipperposition~\cite{Cruanes17}, as these solvers do not support the SMT-LIB input format~\cite{SMTLIB}; yet for further comparison we refer to~\cite{CICM20,CADE21,FMCAD21}.

We ran our experiments using (i) benchmarks over inductive data types (UFDT set of the SMT-LIB benchmark library and {\it dty} set of the inductive benchmarks~\cite{CICM21}), (ii) benchmarks using integers (LIA, UFLIA, NIA and UFNIA of SMT-LIB and {\it int} of~\cite{CICM21}),
and (iii) benchmarks using both integers and data types (UFDTLIA of SMT-LIB). 
From these datasets, we excluded those problems that are  marked satisfiable, as our work is meant for validity checking\footnote{we have excluded all together 1562 satisfiable problems from
LIA, UFLIA, NIA and UFNIA; and  86 satisfiable problems from UFDT.}

\begin{sloppypar}
For our experiments, we used 
\solver{Z3} version 4.8.12 in the default configuration, and \cvc4 version~1.8 with parameters \texttt{--conjecture-gen --quant-ind}.
To extensively compare \vampire{} and \vampire{}*, we ran multiple instances of both for each experiment: we used a portfolio of 18 base configurations differing in the parameters not related to induction.
Additionally, we varied the induction parameters of \vampire{}* for each experiment:
for (i) we used \texttt{--induction struct --structural\_induction\_kind one --induction\_gen on -induction\_on\_complex\_terms on}, for (ii) \texttt{--induction int --induction\_multiclause off},
for (iii) \texttt{--induction both --structural\_induction\_kind one --induction\_gen on -induction\_on\_complex\_terms on}.
In experiments (ii) and (iii), for each of the 18 base configurations we ran 7 instances of \vampire{}* with different integer induction parameters, chosen based on preliminary experimentation on a smaller set of benchmarks.
%
\ifbool{shortversion}{}{
For an overview of the \vampire{}* configurations, see Appendix~\ref{sec:appendix2}.}%
Each prover configuration was given 10 seconds and 16 GB of memory per each problem.
The experiments were ran on computers with 32 cores (AMD Epyc 7502, 2.5 GHz) and 1 TB RAM.
\end{sloppypar}



\subsection{Experimental Results}
\begin{table}[t]
\setlength{\tabcolsep}{0.007\textwidth}
\centering
\begin{tabular}{c||c|c|c|c|c|c|c|c||c}
\hline 
    Problem     & \multicolumn{6}{c|}{SMT-LIB} & \multicolumn{2}{c||}{ind. set~\cite{CICM21}} & \\ \cline{2-9}
    set         & UFDT      & UFDTLIA   & LIA       & UFLIA     & NIA       & UFNIA     & {\it dty} & {\it int} & sum       \\ \hline
    Total count & 4483          & 327       & 404       & 10118     & 8         & 12181     & 3397      & 120       & 31038     \\ \hline\hline
    \vampire{}  & 1848      & 82        & 241       & 6125      & 3         & 3704      & 17        & 0         & 12020     \\ \hline
    \vampire{}* & 1792      & 186       & 241       & 6240      & 4         & 3679      & 464       & 76        & 12682     \\ \hline
    \cvc4       & 2072      & 200       & 357       & 6911      & 7         & 3022      & 164       & 30        & 12763     \\ \hline
    \solver{Z3} & 1807      & 76        & 242       & 6710      & 2         & 4938      & 17        & 0         & 13792     \\
\end{tabular}
\vspace*{0.2em}
\caption{Comparison of the number of solved problems. The configuration of \vampire{} and \vampire{}* depends on the benchmark set.}
\label{tab:results-overall}
\end{table}

\paragraph{Results overview.}
Our results are summarized  in Table~\ref{tab:results-overall}.
For \vampire{} and \vampire{}* we show the number of problems solved by the most successful configuration.
Note that for different benchmark sets the most successful configurations might be different. 
In the inductive problems, 
the maximum and average numbers of induction steps in a proof were 20 and 1.54, respectively, and the maximum number of nested induction steps was 9. 
Overall, Table~\ref{tab:results-overall} shows that  \vampire{}* outperforms \vampire{} without induction. Moreover,  \vampire{}* is competitive with  leading SMT solvers. 

\paragraph{Comparison of \vampire{} and \vampire{}*.}
To evaluate the impact of inductive reasoning in \vampire{},  we look at two key metrics: the \emph{overall number of solved problems}; and the \emph{number of newly solved problems}, which we define as the number of problems solved using induction\footnote{New rules  change proof search organization and  \vampire{}* might solve a problem without using induction, while this problem was not solved by \vampire{}. We do not consider such problems to be newly solved.} by some \vampire{}*, but not solved by any \vampire{}.
The latter metric is especially important, since in practice, one can run multiple solvers or configurations in parallel, and thereby solve the union of all problems solved by individual solvers.

 Table~\ref{tab:results-int-configs} summarizes our result. 
Column ``Combined''  lists the number of problems solved by any instance of the configuration, and in the parentheses the number of problems newly solved by the  configuration. 
The other columns (most solved, most new, default mode) give the numbers of solved problems, and in parentheses newly solved problems,  for the corresponding \vampire{}/\vampire{}* instance.
The ``Default mode'' columns shows results for the best induction configuration with all non-induction parameters set to default.

Induction helped most with the {\tt dty}, {\tt int} and UFDTLIA benchmark sets, as these sets contain a lot of problems focused on induction (induction was used in 91\% of proofs for problems in {\tt dty}, in all proofs in {\tt int}, and in 71\% of proofs in UFDTLIA), while the other sets contain a wide variety of problems (induction was only used in 2\% of proofs in UFDT and 8.8\% of proofs in LIA, UFLIA, NIA and UFNIA).
Interestingly, the configuration which solved most problems in {\tt int} solved the least in LIA, UFLIA, NIA, UFNIA combined, what illustrates the difficulty in choosing the right values for integer induction parameters for such a mixed benchmark set.


\begin{table}[tb]
\setlength{\tabcolsep}{0.007\textwidth}
\ifbool{shortversion}{
 \newcommand\anyvampire{\vampire{}*}
}{ 
 \newcommand\anyvampire{any \vampire{}*}
} 
\newcommand\myline{\cline{2-6}}
\centering
\begin{tabular}{c|c||c|c|c|c}
\hline 
  Benchmarks  &    Configurations   & Combined  & Most solved   & Most new   & Default mode  \\ \hline\hline
\multirowcell{2}{UFDT} & \vampire{} & 2082  & 1848 & -     & 1827      \\ \myline
    & \anyvampire & 2047    &  1792 (12) & 1754 (17)  & 1761      \\ \hline
\multirowcell{2}{{\it dty}}    &  \vampire{} & 17  & 17 & -     &  17     \\ \myline
    & \anyvampire & 525 & 464 (453) & 464 (453) & 432       \\ \hline
\multirowcell{2}{LIA, UFLIA, \\ NIA, UFNIA}    &    \vampire{}      & 11260     & 10073         & -             & 9835          \\ \myline
    & \anyvampire     & 11334 (81) & 10051 (0)    & 9006 (41)     & 9773 (0)      \\ \hline
\multirowcell{2}{{\it int}}    &  \vampire{} & 0  & 0 & -     & 0      \\ \myline
    & \anyvampire & 118 (118) & 76 (76) & 76 (76) & 49 (49)       \\ \hline
\multirowcell{2}{UFDTLIA} & \vampire{} & 91  & 82 & -     & 65      \\ \myline
    & \anyvampire & 197 (108)    & 186 (101)     & 186 (101)  & 136 (72)      \\ \hline
\end{tabular}
\vspace*{0.2em}
\caption{Comparison of  \vampire and \vampire{}* configurations;  numbers given  (in parentheses) indicate new problems solved using induction but not 
without induction.
}
\label{tab:results-int-configs}
\end{table}

\section{Conclusion}
Motivated by application of program analysis and verification, we describe recent advances in automating inductive reasoning 
about first-order (program) properties using inductively defined data types and beyond. We integrate induction in the 
saturation-based proof engine of first-order theorem provers, without radical changes in
the existing machinery of such provers. 
Our inductive inference rules and heuristics open up new research directions to be further studied in automating induction. Guiding and further extending the application of multi-clause induction with theory-specific induction schema variants is an interesting line of research. Combining induction schemas and rules and using lemma generation and rewriting procedures from inductive theorem provers are another ways to further improve saturation-based inductive reasoning. \\

\noindent\textbf{Acknowledgements.} We thank Johannes Schoisswohl for joint work related on experimenting with inductive theorem provers. This work was partially funded by the ERC CoG ARTIST
101002685,  the EPSRC
grant EP/P03408X/1, the FWF grant LogiCS W1255-N23, the
Amazon ARA 2020 award FOREST and the TU Wien SecInt
DK.
%
%
%
\bibliographystyle{splncs04}
\bibliography{bibliography}

\ifbool{shortversion}{}{
\appendix
\newpage
\section{Appendix}\label{sec:appendix}
\ifbool{shortversion}{
}{ 
\subsection{Full proof of Example \ref{eq:ex1}}\label{sec:appendix1}
{\footnotesize
\begin{longtable}{l l r}
    1.& $\forall x.\add(\zero,x)=x$&[axiom]\\
    2.& $\forall x,y.\add(\suc(x), y) = \suc(\add(x,y)))$ &[axiom]\\
    3.& $\even(\zero)$ &[axiom]\\
    4.& $\forall x.\even(\suc(x)) \leftrightarrow \neg\even(x))$ &[axiom]\\
    5.& $\even(\suc(x)) \lor \even(x))$ &[cnf 4]\\
    6.& $\neg\even(\suc(x)) \lor \neg\even(x))$ &[cnf 4]\\
    7.& $\half(\zero) = \zero$ &[axiom]\\
    8.& $\half(\suc(\zero)) = \zero$ &[axiom]\\
    9.& $\forall x.\half(\suc(\suc(x))) = \suc(\half(x)))$ &[axiom]\\
    10.& $\forall x.\even(x) \rightarrow x=\add(\half(x),\half(x)))$ &[conjecture]\\
    11.& $\even(\sigma_1)$ &[cnf 10]\\
    12.& $x\neq\add(\half(\sigma_1),\half(\sigma_1)))$ &[cnf 10]\\
    13. & $\even(\zero)\lor\even(\suc(\zero))\lor\neg\even(\sigma_2)\lor\sigma_2=\add(\half(\sigma_2),\half(\sigma_2))$&[IndMC \\
        & $\quad\lor\neg\even(x)\lor x=\add(\half(x),\half(x))$&11,12]\\
    14. & $\even(\zero)\lor\even(\suc(\zero))\lor\even(\suc(\suc(\sigma_2)))$&[IndMC \\
        & $\quad\lor\neg\even(x)\lor x=\add(\half(x),\half(x))$&11,12]\\
    15. & $\even(\zero)\lor\even(\suc(\zero))$&\\
        & $\quad\lor\suc(\suc(\sigma_2))\neq\add(\half(\suc(\suc(\sigma_2))),\half(\suc(\suc(\sigma_2))))$&[IndMC \\
        & $\quad\lor\neg\even(x)\lor x=\add(\half(x),\half(x))$&11,12]\\
    16. & $\even(\zero)\lor\suc(\zero)\neq\add(\half(\suc(\zero),\suc(\zero)))$&\\
        & $\quad\lor\neg\even(\sigma_2)\lor\sigma_2=\add(\half(\sigma_2),\half(\sigma_2))$&[IndMC \\
        & $\quad\lor\neg\even(x)\lor x=\add(\half(x),\half(x))$&11,12]\\
    17. & $\even(\zero)\lor\suc(\zero)\neq\add(\half(\suc(\zero),\suc(\zero)))$&[IndMC \\
        & $\quad\lor\even(\suc(\suc(\sigma_2)))\lor\neg\even(x)\lor x=\add(\half(x),\half(x))$&11,12]\\
    18. & $\even(\zero)\lor\suc(\zero)\neq\add(\half(\suc(\zero),\suc(\zero)))$&\\
        & $\quad\lor\suc(\suc(\sigma_2))\neq\add(\half(\suc(\suc(\sigma_2))),\half(\suc(\suc(\sigma_2))))$&[IndMC\\
        & $\quad\lor\neg\even(x)\lor x=\add(\half(x),\half(x))$&11,12]\\
    19. & $\zero\neq\add(\half(\zero),\half(\zero))\lor\even(\suc(\zero))$&\\
        & $\quad\lor\neg\even(\sigma_2)\lor\sigma_2=\add(\half(\sigma_2),\half(\sigma_2))$&[IndMC\\
        & $\quad\lor\neg\even(x)\lor x=\add(\half(x),\half(x))$&11,12]\\
    20. & $\zero\neq\add(\half(\zero),\half(\zero))\lor\even(\suc(\zero))$&[IndMC\\
        & $\quad\lor\even(\suc(\suc(\sigma_2)))\lor\neg\even(x)\lor x=\add(\half(x),\half(x))$&11,12]\\
    21. & $\zero\neq\add(\half(\zero),\half(\zero))\lor\even(\suc(\zero))$&\\ 
        & $\quad\lor\suc(\suc(\sigma_2))\neq\add(\half(\suc(\suc(\sigma_2))),\half(\suc(\suc(\sigma_2))))$&[IndMC\\
        & $\quad\lor\neg\even(x)\lor x=\add(\half(x),\half(x))$&11,12]\\
    22. & $\zero\neq\add(\half(\zero),\half(\zero))\lor\suc(\zero)\neq\add(\half(\suc(\zero),\suc(\zero)))$&\\
        & $\quad\lor\neg\even(\sigma_2)\lor\sigma_2=\add(\half(\sigma_2),\half(\sigma_2))$&[IndMC\\
        & $\quad\lor\neg\even(x)\lor x=\add(\half(x),\half(x))$&11,12]\\
    23. & $\zero\neq\add(\half(\zero),\half(\zero))\lor\suc(\zero)\neq\add(\half(\suc(\zero),\suc(\zero)))$&[IndMC\\
        & $\quad\lor\even(\suc(\suc(\sigma_2)))\lor\neg\even(x)\lor x=\add(\half(x),\half(x))$&11,12]\\
    24. & $\zero\neq\add(\half(\zero),\half(\zero))\lor\suc(\zero)\neq\add(\half(\suc(\zero),\suc(\zero)))$\\
        & $\quad\lor\suc(\suc(\sigma_2))\neq\add(\half(\suc(\suc(\sigma_2))),\half(\suc(\suc(\sigma_2))))$&[IndMC \\ 
        & $\quad\lor\neg\even(x)\lor x=\add(\half(x),\half(x))$&11,12]\\
    25. & $\even(\zero)\lor\even(\suc(\zero))\lor\neg\even(\sigma_2)\lor$&\\ 
        & $\quad\sigma_2=\add(\half(\sigma_2),\half(\sigma_2))\lor\neg\even(\sigma_1)$&[BR 13,12]\\
    26. & $\even(\zero)\lor\even(\suc(\zero))\lor\neg\even(\sigma_2)\lor\sigma_2=\add(\half(\sigma_2),\half(\sigma_2))$&[BR 25,11]\\
    27. & $\even(\zero)\lor\even(\suc(\zero))\lor\even(\suc(\suc(\sigma_2)))\lor\neg\even(\sigma_1)$&[BR 14,12]\\
    28. & $\even(\zero)\lor\even(\suc(\zero))\lor\even(\suc(\suc(\sigma_2)))$&[BR 27,11]\\
    29. & $\even(\zero)\lor\even(\suc(\zero))$&\\ 
        & $\quad\lor\suc(\suc(\sigma_2))\neq\add(\half(\suc(\suc(\sigma_2))),\half(\suc(\suc(\sigma_2))))\lor\neg\even(\sigma_1)$&[BR 15,12]\\
    30. & $\even(\zero)\lor\even(\suc(\zero))$&\\
        & $\quad\lor\suc(\suc(\sigma_2))\neq\add(\half(\suc(\suc(\sigma_2))),\half(\suc(\suc(\sigma_2))))$&[BR 29,11]\\
    31. & $\even(\zero)\lor\suc(\zero)\neq\add(\half(\suc(\zero),\suc(\zero)))\lor\neg\even(\sigma_2)$\\
        & $\quad\lor\sigma_2=\add(\half(\sigma_2),\half(\sigma_2))\lor\neg\even(\sigma_1)$&[BR 16,12]\\
    32. & $\even(\zero)\lor\suc(\zero)\neq\add(\half(\suc(\zero),\suc(\zero)))\lor\neg\even(\sigma_2)$\\
        & $\quad\lor\sigma_2=\add(\half(\sigma_2),\half(\sigma_2))$&[BR 31,11]\\
    33. & $\even(\zero)\lor\suc(\zero)\neq\add(\half(\suc(\zero),\suc(\zero)))$&\\
        & $\quad\lor\even(\suc(\suc(\sigma_2)))\lor\neg\even(\sigma_1)$&[BR 17,12]\\
    34. & $\even(\zero)\lor\suc(\zero)\neq\add(\half(\suc(\zero),\suc(\zero)))\lor\even(\suc(\suc(\sigma_2)))$&[BR 33,11]\\
    35. & $\even(\zero)\lor\suc(\zero)\neq\add(\half(\suc(\zero),\suc(\zero)))$&\\ 
        & $\quad\lor\suc(\suc(\sigma_2))\neq\add(\half(\suc(\suc(\sigma_2))),\half(\suc(\suc(\sigma_2))))\lor\neg\even(\sigma_1)$&[BR 18,12]\\
    36. & $\even(\zero)\lor\suc(\zero)\neq\add(\half(\suc(\zero),\suc(\zero)))$&\\ 
        & $\quad\lor\suc(\suc(\sigma_2))\neq\add(\half(\suc(\suc(\sigma_2))),\half(\suc(\suc(\sigma_2))))$&[BR 35,11]\\
    37. & $\zero\neq\add(\half(\zero),\half(\zero))\lor\even(\suc(\zero))\lor\neg\even(\sigma_2)$&\\ 
        & $\quad\lor\sigma_2=\add(\half(\sigma_2),\half(\sigma_2))\lor\neg\even(\sigma_1)$&[BR 19,12]\\
    38. & $\zero\neq\add(\half(\zero),\half(\zero))\lor\even(\suc(\zero))\lor\neg\even(\sigma_2)$&\\ 
        & $\quad\lor\sigma_2=\add(\half(\sigma_2),\half(\sigma_2))$&[BR 37,11]\\
    39. & $\zero\neq\add(\half(\zero),\half(\zero))\lor\even(\suc(\zero))$&\\
        & $\quad\lor\even(\suc(\suc(\sigma_2)))\lor\neg\even(\sigma_1)$&[BR 20,12]\\
    40. & $\zero\neq\add(\half(\zero),\half(\zero))\lor\even(\suc(\zero))\lor\even(\suc(\suc(\sigma_2)))$&[BR 39,11]\\
    41. & $\zero\neq\add(\half(\zero),\half(\zero))\lor\even(\suc(\zero))$&\\ 
        & $\quad\lor\suc(\suc(\sigma_2))\neq\add(\half(\suc(\suc(\sigma_2))),\half(\suc(\suc(\sigma_2))))\lor\neg\even(\sigma_1)$&[BR 21,12]\\
    42. & $\zero\neq\add(\half(\zero),\half(\zero))\lor\even(\suc(\zero))$&\\ 
        & $\quad\lor\suc(\suc(\sigma_2))\neq\add(\half(\suc(\suc(\sigma_2))),\half(\suc(\suc(\sigma_2))))$&[BR 41,11]\\
    43. & $\zero\neq\add(\half(\zero),\half(\zero))\lor\suc(\zero)\neq\add(\half(\suc(\zero),\suc(\zero)))$&\\
        & $\quad\lor\neg\even(\sigma_2)\lor\sigma_2=\add(\half(\sigma_2),\half(\sigma_2))\lor\neg\even(\sigma_1)$&[BR 22,12]\\
    44. & $\zero\neq\add(\half(\zero),\half(\zero))\lor\suc(\zero)\neq\add(\half(\suc(\zero),\suc(\zero)))$&\\
        & $\quad\lor\neg\even(\sigma_2)\lor\sigma_2=\add(\half(\sigma_2),\half(\sigma_2))$&[BR 43,11]\\
    45. & $\zero\neq\add(\half(\zero),\half(\zero))\lor\suc(\zero)\neq\add(\half(\suc(\zero),\suc(\zero)))$&\\
        & $\quad\lor\even(\suc(\suc(\sigma_2)))\lor\neg\even(\sigma_1)$&[BR 23,12]\\
    46. & $\zero\neq\add(\half(\zero),\half(\zero))\lor\suc(\zero)\neq\add(\half(\suc(\zero),\suc(\zero)))$&\\
        & $\quad\lor\even(\suc(\suc(\sigma_2)))$&[BR 45,11]\\
    47. & $\zero\neq\add(\half(\zero),\half(\zero))\lor\suc(\zero)\neq\add(\half(\suc(\zero),\suc(\zero)))$\\
        & $\quad\lor\suc(\suc(\sigma_2))\neq\add(\half(\suc(\suc(\sigma_2))),\half(\suc(\suc(\sigma_2))))$&\\ 
        & $\quad\lor\neg\even(x)\lor x=\add(\half(x),\half(x))$&[BR 24,12]\\
    48. & $\zero\neq\add(\half(\zero),\half(\zero))\lor\suc(\zero)\neq\add(\half(\suc(\zero),\suc(\zero)))$\\
        & $\quad\lor\suc(\suc(\sigma_2))\neq\add(\half(\suc(\suc(\sigma_2))),\half(\suc(\suc(\sigma_2))))$&\\ 
        & $\quad\lor\neg\even(x)\lor x=\add(\half(x),\half(x))$&[BR 47,11]\\
    49. & $\zero\neq\add(\zero,\zero)\lor\even(\suc(\zero))\lor\neg\even(\sigma_2)$&\\
        & $\quad\lor\sigma_2=\add(\half(\sigma_2),\half(\sigma_2))$&[Sup 38,7]\\
    50. & $\zero\neq\zero\lor\even(\suc(\zero))\lor\neg\even(\sigma_2)\lor\sigma_2=\add(\half(\sigma_2),\half(\sigma_2))$&[Sup 49,1]\\
    51. & $\even(\suc(\zero))\lor\neg\even(\sigma_2)\lor\sigma_2=\add(\half(\sigma_2),\half(\sigma_2))$&[ER 50]\\
    52. & $\neg\even(\zero)\lor\neg\even(\sigma_2)\lor\sigma_2=\add(\half(\sigma_2),\half(\sigma_2))$&[BR 51,6]\\
    53. & $\neg\even(\sigma_2)\lor\sigma_2=\add(\half(\sigma_2),\half(\sigma_2))$&[BR 52,3]\\
    54. & $\zero\neq\add(\zero,\zero)\lor\even(\suc(\zero))\lor\even(\suc(\suc(\sigma_2)))$&[Sup 40,7]\\
    55. & $\zero\neq\zero\lor\even(\suc(\zero))\lor\even(\suc(\suc(\sigma_2)))$&[Sup 54,1]\\
    56. & $\even(\suc(\zero))\lor\even(\suc(\suc(\sigma_2)))$&[ER 55]\\
    57. & $\neg\even(\zero)\lor\even(\suc(\suc(\sigma_2)))$&[BR 56,6]\\
    58. & $\even(\suc(\suc(\sigma_2)))$&[BR 57,3]\\
    59. & $\neg\even(\suc(\sigma_2))$&[BR 58,6]\\
    60. & $\even(\sigma_2)$&[BR 59,5]\\
    61. & $\sigma_2=\add(\half(\sigma_2),\half(\sigma_2))$&[BR 60,53]\\
    62. & $\zero\neq\add(\zero,\zero)\lor\even(\suc(\zero))$&\\ 
        & $\quad\lor\suc(\suc(\sigma_2))\neq\add(\half(\suc(\suc(\sigma_2))),\half(\suc(\suc(\sigma_2))))$&[Sup 42,7]\\
    63. & $\zero\neq\zero\lor\even(\suc(\zero))$&\\
        & $\quad\lor\suc(\suc(\sigma_2))\neq\add(\half(\suc(\suc(\sigma_2))),\half(\suc(\suc(\sigma_2))))$&[Sup 62,1]\\
    64. & $\even(\suc(\zero))\lor\suc(\suc(\sigma_2))\neq\add(\half(\suc(\suc(\sigma_2))),\half(\suc(\suc(\sigma_2))))$&[ER 63]\\
    65. & $\neg\even(\zero)\lor\suc(\suc(\sigma_2))\neq\add(\half(\suc(\suc(\sigma_2))),\half(\suc(\suc(\sigma_2))))$&[BR 64,6]\\
    66. & $\suc(\suc(\sigma_2))\neq\add(\half(\suc(\suc(\sigma_2))),\half(\suc(\suc(\sigma_2))))$&[BR 65,3]\\
    67. & $\suc(\suc(\sigma_2))\neq\add(\suc(\half(\sigma_2)),\suc(\half(\sigma_2)))$&[Sup 66,9]\\
    68. & $\suc(\suc(\sigma_2))\neq\suc(\add(\half(\sigma_2),\suc(\half(\sigma_2))))$&[Sup 67,2]\\
    69. & $\suc(\sigma_2)\neq\add(\half(\sigma_2),\suc(\half(\sigma_2)))$&[inj $\suc$ 68]\\

    70. & $\suc(\add(\zero,\half(\sigma_2)))\neq\add(\zero,\suc(\half(\sigma_2)))$&\\ 
        & $\quad\lor\suc(\add(\sigma_3,\half(\sigma_2)))=\add(\sigma_3,\suc(\half(\sigma_2)))$&[IndHRW \\
        & $\quad\lor\suc(\add(x,\half(\sigma_2)))=\add(x,\suc(\half(\sigma_2)))$ &69,61]\\
    71. & $\suc(\add(\zero,\half(\sigma_2)))\neq\add(\zero,\suc(\half(\sigma_2)))$&\\ 
        & $\quad\lor\suc(\add(\suc(\sigma_3),\half(\sigma_2)))\neq\add(\suc(\sigma_3),\suc(\half(\sigma_2)))$&[IndHRW \\
        & $\quad\lor\suc(\add(x,\half(\sigma_2)))=\add(x,\suc(\half(\sigma_2)))$ &69,61]\\
    72. & $\suc(\add(\zero,\half(\sigma_2)))\neq\add(\zero,\suc(\half(\sigma_2)))$&\\ 
        & $\quad\lor\suc(\add(\sigma_3,\half(\sigma_2)))=\add(\sigma_3,\suc(\half(\sigma_2)))$&\\
        & $\quad\lor\suc(\sigma_2)=\add(\half(\sigma_2),\suc(\half(\sigma_2)))$ &[Sup 70,61]\\
    73. & $\suc(\add(\zero,\half(\sigma_2)))\neq\add(\zero,\suc(\half(\sigma_2)))$&\\ 
        & $\quad\lor\suc(\add(\suc(\sigma_3),\half(\sigma_2)))\neq\add(\suc(\sigma_3),\suc(\half(\sigma_2)))$&\\
        & $\quad\lor\suc(\sigma_2)=\add(\half(\sigma_2),\suc(\half(\sigma_2)))$ &[Sup 71,61]\\
    74. & $\suc(\add(\zero,\half(\sigma_2)))\neq\add(\zero,\suc(\half(\sigma_2)))$&\\ 
        & $\quad\lor\suc(\add(\sigma_3,\half(\sigma_2)))=\add(\sigma_3,\suc(\half(\sigma_2)))$ &[BR 72,69]\\
    75. & $\suc(\add(\zero,\half(\sigma_2)))\neq\add(\zero,\suc(\half(\sigma_2)))$&\\ 
        & $\quad\lor\suc(\add(\suc(\sigma_3),\half(\sigma_2)))\neq\add(\suc(\sigma_3),\suc(\half(\sigma_2)))$ &[BR 73,69]\\
    76. & $\suc(\half(\sigma_2))\neq\suc(\half(\sigma_2))$&\\ 
        & $\quad\lor\suc(\add(\sigma_3,\half(\sigma_2)))=\add(\sigma_3,\suc(\half(\sigma_2)))$ &[Sup 74,1]\\
    77. & $\suc(\add(\sigma_3,\half(\sigma_2)))=\add(\sigma_3,\suc(\half(\sigma_2)))$ &[ER 76]\\
    78. & $\suc(\half(\sigma_2))\neq\suc(\half(\sigma_2))$&\\ 
        & $\quad\lor\suc(\add(\suc(\sigma_3),\half(\sigma_2)))\neq\add(\suc(\sigma_3),\suc(\half(\sigma_2)))$ &[Sup 75,1]\\
    79. & $\suc(\add(\suc(\sigma_3),\half(\sigma_2)))\neq\add(\suc(\sigma_3),\suc(\half(\sigma_2)))$ &[ER 78]\\
    80. & $\suc(\suc(\add(\sigma_3,\half(\sigma_2))))\neq\add(\suc(\sigma_3),\suc(\half(\sigma_2)))$ &[Sup 79,2]\\
    81. & $\suc(\suc(\add(\sigma_3,\half(\sigma_2))))\neq\suc(\add(\sigma_3,\suc(\half(\sigma_2))))$ &[Sup 80,2]\\
    82. & $\suc(\add(\sigma_3,\half(\sigma_2)))\neq\add(\sigma_3,\suc(\half(\sigma_2)))$ &[inj $\suc$ 81]\\
    83. & $\qed$ &[BR 82,77]
\end{longtable}
}

\subsection{\vampire{} configurations used in experiments}
} 
\label{sec:appendix2}
\renewcommand{\floatpagefraction}{.8}%
\begin{table}
\setlength{\tabcolsep}{0.007\textwidth}
\begin{tabular}{p{3.7cm}||p{0.5cm}|p{1.65cm}|p{1.65cm}|p{1.65cm}|p{0.5cm}|p{1.25cm}}
    & \multicolumn{6}{c}{Value used in configuration with ID:} \\ \cline{2-7}
  Parameter                             & 0     & 1     & 2     & 3         & 4     & 5     \\ \hline\hline
  \texttt{--age\_weight\_ratio}         & 1     & 1     & 2     & 3         & 5     & 10    \\ \hline
  \texttt{--saturation\_algorithm}      & lrs   & lrs   & lrs   & discount  & lrs   & discount \\\hline
  \texttt{--selection}                  & 10    & 11    & 1010  & 11        & 4     & 1011        \\ \hline
  \texttt{--theory\_instantiation}      & off   & off   & off   & strong    & off   & off    \\\hline
  \texttt{--unification\_with\_}\newline \phantom{\texttt{--}}\texttt{abstraction} & off & one\_side\_\newline interpreted & one\_side\_\newline interpreted & one\_side\_\newline interpreted & off & off  \\\hline
\end{tabular} \vspace*{1em} \\
\setlength{\tabcolsep}{0.007\textwidth}
\begin{tabular}{p{6.2cm}||p{1.65cm}|p{1.65cm}|p{1.9cm}}
    \hline
    & \multicolumn{3}{c}{Value used in configuration with ID:} \\ \cline{2-4}
  Parameter                                 & A     & B     & C   \\ \hline\hline
  \texttt{--term\_ordering}                 & kbo   & lpo   & lpo      \\ \hline
  \texttt{--demodulation\_redundancy\_check} & on   & off   & off   \\\hline
  \texttt{--unit\_resulting\_resolution}    & off   & off   & on         \\ \hline
  \texttt{--sos}                            & off   & theory& off     \\\hline
  \texttt{--sos\_theory\_limit}             & 0     & 1     & 0   \\\hline
  \texttt{--evaluation}                     & simple & simple & force     \\\hline
  \texttt{--gaussian\_variable\_elimination} & off   & off  & force     \\\hline
  \texttt{--arithmetic\_subterm\_generalizations} & off & off & force     \\\hline
  \texttt{--push\_unary\_minus}             & off   & off   & on     \\\hline
  \texttt{--cancellation}                   & off   & off   & force   
\end{tabular}
\vspace*{0.5em}
\caption{General parameter combinations for configurations with ID \{0-5\}\{A-C\}}\label{tab:appendix-configs1}
\end{table}
\begin{table}
\resizebox{\textwidth}{!}{
\begin{tabular}{p{2.6cm}||p{1.55cm}|p{1.55cm}|p{1.55cm}|p{0.95cm}|p{1.55cm}|p{1.55cm}|p{1.1cm}}
    & \multicolumn{7}{c}{Value used in configuration with ID:} \\ \cline{2-8}
  Parameter                                 & 322c0 & 031c1 & 121c0 & 030-1 & 011-1 & 001-1 & 140-0d     \\ \hline\hline
  \texttt{--int\_induction\_\newline strictness\_eq}   & not\_in\newline \_both   & none   & toplevel\newline \_not\_in\_\newline other   & none   & none   & none   & toplevel\newline \_not\_in\_\newline other \\ \hline
  \texttt{--int\_induction\_\newline strictness\_comp}     & only\_one\_\newline occurrence   & not\_in\newline \_both   & only\_one\_\newline occurrence   & not\_in\newline \_both   & toplevel\newline \_not\_in\_\newline other   & none   & always \\ \hline
  \texttt{--int\_induction\_\newline strictness\_term}     & no\_skolems   & interpreted\newline \_constant   & interpreted\newline \_constant   & none   & interpreted\newline \_constant   & interpreted\newline \_constant   & none \\ \hline
  \texttt{--induction\_on\_\newline complex\_terms}  & on    & on    & on    & off   & off   & off   & off \\ \hline
  \texttt{--induction\_max\_\newline depth}          &  0    & 1     & 0     & 1     & 1     & 1     & 0   \\ \hline
  \texttt{--int\_induction\_\newline default\_bound} & off   & off   & off   & off   & off   & off   & on
\end{tabular}}
\vspace*{0.5em}
\caption{Integer induction parameter combinations for configurations with ID 322c0, 031c1, 121c0, 030-1, 011-1, 001-1 and 140-0d.}\label{tab:appendix-configs2}
\end{table}


The 18 strategies we used as base configurations for both \vampire{} and \vampire{}* consisted of all combinations of configurations 0-5 and A-C displayed in Table~\ref{tab:appendix-configs1}. Configurations 0 and A correspond to the default values.

The 7 integer induction configurations used for \vampire{}* are displayed in Table~\ref{tab:appendix-configs2}.
\ifbool{shortversion}{}{
For the \texttt{--int\_induction\_strictness} parameter values, each digit controls an aspect of integer induction: what equality literals we apply induction on, what comparison literals, what terms. The larger the value the less we apply induction.}
}
%




\end{document}